\documentclass[10pt]{article}
\usepackage{amssymb,amsmath}
\usepackage{amsthm}
\usepackage{mathrsfs}  
\usepackage{graphicx}
\usepackage{dcolumn}
\usepackage{bm}
\usepackage{fullpage}
\usepackage{color}
\usepackage[all]{xy}
\begin{document}
\newtheorem{Def}{Definition}[section]
\newtheorem{Thm}{Theorem}[section]
\newtheorem{Proposition}{Proposition}[section] 
\newtheorem{Lemma}{Lemma}[section]
\theoremstyle{definition}
\newtheorem*{Proof}{Proof}
\newtheorem*{Example}{Example}
\newtheorem{Postulate}{Postulate}[section]
\newtheorem{Corollary}{Corollary}[section]
\newtheorem{Remark}{Remark}[section]
\theoremstyle{remark} 
\newcommand{\beq}{\begin{equation}}
\newcommand{\beqa}{\begin{eqnarray}}
\newcommand{\eeq}{\end{equation}}
\newcommand{\eeqa}{\end{eqnarray}}
\newcommand{\non}{\nonumber}
\newcommand{\fr}[1]{(\ref{#1})}
\newcommand{\cc}{\mbox{c.c.}}
\newcommand{\nr}{\mbox{n.r.}}
\newcommand{\eq}{\mathrm{eq}}
\newcommand{\B}{\mathrm{B}}
\newcommand{\Ising}{\mathrm{Ising}}
\newcommand{\heatbath}{\mathrm{heat\, bath}}
\newcommand{\atan}{\mathrm{Tan}{}^{-1}}
\newcommand{\atanh}{\mathrm{Tanh}{}^{-1}}
\newcommand{\acosh}{\mathrm{Cosh}{}^{-1}}
\newcommand{\bb}{\mbox{\boldmath {$b$}}}
\newcommand{\bbe}{\mbox{\boldmath {$e$}}}
\newcommand{\bt}{\mbox{\boldmath {$t$}}}
\newcommand{\bn}{\mbox{\boldmath {$n$}}}
\newcommand{\br}{\mbox{\boldmath {$r$}}}
\newcommand{\bC}{\mbox{\boldmath {$C$}}}
\newcommand{\bH}{\mbox{\boldmath {$H$}}}
\newcommand{\bp}{\mbox{\boldmath {$p$}}}
\newcommand{\bx}{\mbox{\boldmath {$x$}}}
\newcommand{\bF}{\mbox{\boldmath {$F$}}}
\newcommand{\bT}{\mbox{\boldmath {$T$}}}
\newcommand{\bomega}{\mbox{\boldmath {$\omega$}}}
\newcommand{\ve}{{\varepsilon}}
\newcommand{\e}{\mathrm{e}}
\newcommand{\F}{\mathrm{F}}
\newcommand{\Loc}{\mathrm{Loc}}
\newcommand{\Ree}{\mathrm{Re}}
\newcommand{\Imm}{\mathrm{Im}}
\newcommand{\hF}{\widehat F}
\newcommand{\hL}{\widehat L}
\newcommand{\tA}{\widetilde A}
\newcommand{\tB}{\widetilde B}
\newcommand{\tC}{\widetilde C}
\newcommand{\tL}{\widetilde L}
\newcommand{\tK}{\widetilde K}
\newcommand{\tX}{\widetilde X}
\newcommand{\tY}{\widetilde Y}
\newcommand{\tU}{\widetilde U}
\newcommand{\tZ}{\widetilde Z}
\newcommand{\talpha}{\widetilde \alpha}
\newcommand{\te}{\widetilde e}
\newcommand{\tv}{\widetilde v}
\newcommand{\ts}{\widetilde s}
\newcommand{\tx}{\widetilde x}
\newcommand{\ty}{\widetilde y}
\newcommand{\ud}{\underline{\delta}}
\newcommand{\uD}{\underline{\Delta}}
\newcommand{\chN}{\check{N}}
\newcommand{\cA}{{\cal A}}
\newcommand{\cB}{{\cal B}}
\newcommand{\cC}{{\cal C}}
\newcommand{\cD}{{\cal D}}
\newcommand{\cF}{{\cal F}}
\newcommand{\cI}{{\cal I}}
\newcommand{\cL}{{\cal L}}
\newcommand{\cM}{{\cal M}}
\newcommand{\cN}{{\cal N}}
\newcommand{\cO}{{\cal O}}
\newcommand{\cP}{{\cal P}}
\newcommand{\cR}{{\cal R}}
\newcommand{\cS}{{\cal S}}
\newcommand{\cY}{{\cal Y}}
\newcommand{\cZ}{{\cal Z}}
\newcommand{\cU}{{\cal U}}
\newcommand{\cV}{{\cal V}}
\newcommand{\dr}{\mathrm{d}}
\newcommand{\sech}{\mathrm{sech}}
\newcommand{\Exp}{\mathrm{Exp}}
\newcommand{\inp}[2]{\left\langle\,  #1\, , \, #2\, \right\rangle}
\newcommand{\equp}[1]{\overset{\mathrm{#1}}{=}}
\newcommand{\wt}[1]{\widetilde{#1}}
\newcommand{\wh}[1]{\widehat{#1}}
\newcommand{\ch}[1]{\check{#1}}
\newcommand{\ii}{\imath}
\newcommand{\ic}{\iota}
\newcommand{\scrH}{\mathscr{H}}
\newcommand{\mi}{\,\mathrm{i}\,}
\newcommand{\mr}{\,\mathrm{r}\,}
\newcommand{\mbbC}{\mathbb{C}}
\newcommand{\mbbE}{\mathbb{E}}
\newcommand{\mbbR}{\mathbb{R}}
\newcommand{\mbbZ}{\mathbb{Z}}
\newcommand{\ol}[1]{\overline{#1}}
\newcommand{\rmC}{\mathrm{C}}
\newcommand{\rmH}{\mathrm{H}}
\newcommand{\Id}{\mathrm{Id}} 
\newcommand{\avg}[1]{\left\langle\,{#1}\, \right\rangle}
\newcommand{\Leg}{\mathbb{L}}
\newcommand{\avgg}[1]{\left\langle\langle\,{#1}\, \rangle\right\rangle}
\title{
Information and contact geometric description of  \\
expectation variables exactly derived from master equations
}
\author{  Shin-itiro GOTO$^{1}$ and  Hideitsu HINO$^{2}$,\\ 
1)Department of Applied Mathematics and Physics, 
Graduate School of Informatics, \\
Kyoto University, Yoshida Honmachi, Sakyo-ku, Kyoto, 606-8501, Japan\\
2) The Institute of Statistical Mathematics,\\ 
10-3 Midori-cho, Tachikawa, Tokyo 190-8562, Japan
} 
\date{\today}
\maketitle
\begin{abstract}%
In this paper a class of dynamical systems 
describing expectation variables 
exactly derived from 
continuous-time master equations is introduced and studied from 
the viewpoint of differential geometry, where 
such master equations consist of a set of appropriately chosen 
Markov kernels.
To geometrize such dynamical systems for expectation variables, 
information geometry is used for expressing equilibrium states, 
and contact geometry is used for nonequilibrium states. 
Here time-developments of the expectation variables 
are identified with contact Hamiltonian vector fields on a contact manifold.
Also, it is shown that the convergence rate of this dynamical system 
is exponential.  Duality emphasized in information geometry is also 
addressed throughout. 
\end{abstract}%
\section{Introduction}
Information geometry is a geometrization of mathematical statistics \cite{AN}, 
and its structure and applications have been investigated. 
This geometry offers 
tools to study statistical quantities defined on statistical manifolds,  
where statistical manifolds are identified with parameter spaces 
for parametric distribution functions. 
Examples of applications of information geometry include 
statistical interference, 
quantum information, and thermodynamics \cite{Ay2017,Hayashi2006,Naudts2011}. 
From these examples one sees that 
the application of information geometry to sciences and engineering 
enables one 
to visualize theories and 
to utilize differential geometric tools for their analysis.
Thus, one is interested in how to extend information geometry. 
Since equilibrium thermodynamics can be formulated with information geometry, 
a geometrization of nonequilibrium  
thermodynamics is one of the keys for this task.

Contact geometry is known to be an 
odd-dimensional counterpart of symplectic geometry \cite{Arnold,Silva2008}. 
Along with the context stated above,  
one extension of information geometry 
is to use contact geometry \cite{Goto2015,Goto2016}, 
This extension is 
consistent with 
geometrization of equilibrium 
thermodynamics \cite{Mrugala1978,Mrugala2000,Bravetti2015,Grmela2014,Rajeev2008,
Bravetti2018}. 
In Ref.\,\cite{Goto2015}, a  contact geometric description of 
relaxation processes in nonequilibrium thermodynamic systems 
was proposed, and a link between information geometry and contact geometry 
was clarified. 
Since Markov chains can be used as models 
of nonequilibrium thermodynamic systems \cite{Glauber1967}, 
it can be expected that  
contact geometry and Markov chains are related. 
From this, 
master equations and  
Markov Chain Monte Carlo (MCMC)  
methods are expected to be formulated in terms of  contact and 
information geometries. 

Continuous-time master equations are first order ordinary 
differential equations. These and Fokker-Plank equations  
have been  used to describe 
the time-developments of distribution functions 
\cite{Kampen2007,Gardiner2009}. 
Master equations are closely related to 
MCMC methods \cite{Girolami2013}  
and have been used to model nonequilibrium statistical 
phenomena \cite{Glauber1967}. 
Further development of theory of master equations 
is expected to yield those of 
MCMC methods and nonequilibrium statistical physics.  
MCMC methods 
 are well-known methods for obtaining expectation values (averaged values)  
 of some quantities with respect to target distribution functions numerically. 
 These methods have been applied in various disciplines  
 including mathematical engineering, physics, statistics  
 and so on \cite{Metropolis1953,Kirkpatrick1983,Iba2001,Parisi1992, Lyubartsev1992,Hukushima1996}.
For constructing MCMC methods, master equations with their  
Markov kernels play fundamental roles. 
Also it should be noted that dynamical systems on statistical manifolds without
contact geometry   
have been studied in the literature 
\cite{Fujiwara-Amari1995, Nakamura1994,Uwano2016,Noda2011}. 
In addition, information geometric descriptions for Markov chains 
were investigated in the literature as well
\cite{Takabatake2004,Nagaoka2005,Nielsen2018}.

In this paper a class of dynamical systems 
describing expectation variables  
exactly derived from 
continuous-time master equations is studied from the viewpoints  
of  information geometry and contact geometry, where 
such master equations consist of a set of appropriately chosen 
Markov kernels. 
To formulate this class of dynamical systems  
with contact geometry,  
configuration 
space is identified with a contact manifold, and dynamics 
is described as contact Hamiltonian dynamical systems.
It is then shown that the time-asymptotic limit of the 
expectation variables defined in closed dynamical systems  
is consistent with information geometry. 
Also, convergence rates 
are explicitly  calculated after introducing a metric 
tensor field.
Since the present study is closely related to Theorem 4.3 in 
Ref.\,\cite{Goto2015}, the main contributions of  
the present study from the previous one  are stated here.
In this paper, (i) a general class of master equations is 
introduced such that 
a particular example used in Ref.\,\cite{Goto2015} is included,  
(ii) a nonequilibrium free-energy is introduced, (iii) 
a convergence rate is calculated, and (iv) duality in the sense of 
that used in information is stated even in nonequilibrium states.

Some new terminologies will appear in this paper, and they
play various roles.  
The relations among them are briefly listed below. 
\begin{table}[htb]
\begin{center}
\caption{Relations among introduced systems, equations, and methods}
\begin{tabular}{|c||c|c|}
\hline
Sys. for distribution func.&Sys. for observables& Geometric descriptions of observables\\
\hline\hline
Primary master Eq.&Primary moment dynamical Sys.&Contact Hamiltonian Sys. in 
Prop.\,\ref{fact-moment-dynamics-contact-Hamiltonian-system}\\
\hline
Dual master Eq.&Dual moment dynamical Sys.&Contact Hamiltonian Sys. in 
Prop.\,\ref{fact-parameter-dynamics-contact-Hamiltonian-system}\\
\hline
\end{tabular}
\end{center}
\end{table}

This paper is organized as follows. 
In Section\,\ref{section-solvable-heat-bath}, 
a set of master equations is introduced by choosing Markov kernels,  
and such equations  
are termed  primary 
master equations.  
After dual master equations  
are introduced, the explicit 
solutions of these equations are shown.  
In Section\,\ref{section-time-development-observables},
with the use of the clarified features of 
the primary and dual master equations,  
differential equations describing time-development of 
some observables are derived. 
In Section\,\ref{section-geometry-solvable-extended-heat-bath}, 
a geometrization of discussions in Section\,\ref{section-time-development-observables}
is given. Finally, Section\,\ref{section-conclusion} summarizes 
this paper and discusses some future works.

\section{Solvable master equations}
\label{section-solvable-heat-bath}
In this section 
a set of master equations with particular Markov kernels
is introduced, 
and then its solvability is shown first. 
In arguing  this, it is shown how master equations play roles.   
Then, the idea called ``dual'' used in information geometry is 
imported to the master equations. 
With this idea, dual master equation is introduced. 
To emphasize this duality, master equation is  
also referred to as primary master equation.

\subsection{Primary master equation}
In the following a class of master equations is introduced,  
and its solvable feature is explicitly clarified. 
After this, primary master equation is defined.

Let $\Gamma$ be a set of finite discrete states,  $t\in\mbbR$ time, and 
$p(j,t)\,\dr t$ a probability that a state $j\in \Gamma$ is found in between 
$t$ and $t+\dr t$. 
The first objective is to realize a given distribution function 
$p_{\,\theta}^{\,\eq}$ 
that can be written as 
$$
p_{\,\theta}^{\,\eq}(j)
=\frac{\pi_{\,\theta}(j)}{Z(\theta)},
$$
where $\theta\in\Theta\subset\mbbR^{\,n}$ is a parameter set with 
$\theta=\{\theta^{\,1},\ldots,\theta^{\,n}\}$, and 
$Z :\Theta\to\mbbR$  the so-called a partition function so that 
$p_{\,\theta}^{\,\eq}$ is normalized. Thus,  
$$
Z(\theta)
=\sum_{j\in \Gamma}\pi_{\theta}(j),\quad\mbox{so that}\quad
\sum_{j\in \Gamma}p_{\,\theta}^{\,\eq}(j)
=1.
$$

One way to achieve the objective is to employ a 
special set of Markov kernels. 
In what follows, 
how this objective is done is shown. 

Let $p:\Gamma\times\mbbR\to \mbbR_{\geq 0}$  
be a time-dependent probability function. 
Then, consider the set of master equations 
\beq
\frac{\partial}{\partial t}p(j,t)
=\sum_{j'(\neq j)}\left[\,
w(j|j^{\,\prime})\,p(j',t)-w(j^{\,\prime}|j)\,p(j,t)
\,\right],
\label{master-equation-general}
\eeq
where $w:\Gamma\times\Gamma\to I$, ($I:=[\,0,1\,]\subset\mbbR$)  
is such that    
$w(j|j^{\,\prime})\in I$ denotes a probability that a state jumps from 
$j^{\,\prime}$ to $j$.

Assume the following.
\begin{itemize}
\item
Choose $w$ to be 
$$
w_{\,\theta}(j|j^{\,\prime})
=p_{\,\theta}^{\,\eq}(j).
$$
\item
The target distribution $p_{\,\theta}^{\,\eq}(j)$ for any state $j\in\Gamma$ 
does not vanish: $p_{\,\theta}^{\,\eq}(j)\neq 0$.
\end{itemize}

 To show an explicit form of 
$p(j,t)$, one rewrites \fr{master-equation-general} with the 
assumptions.
The equation of motion for $p$ is derived from 
$$
\sum_{j^{\,\prime}(\neq j)}p(j^{\,\prime},t)
=1-p(j,t),
$$
as 
\beq
\frac{\partial}{\partial t}p(j,t)
=\sum_{j'(\neq j)}\left[\,
w_{\,\theta}(j|j^{\,\prime})\,p(j',t)-w_{\,\theta}(j^{\,\prime}|j)\,p(j,t)
\,\right]
=p_{\,\theta}^{\,\eq}(j)-p(j,t).
\label{master-equation-solvable}
\eeq

In this paper \fr{master-equation-solvable} is termed. 
\begin{Def}
(Primary master equations). The set of equations   
\fr{master-equation-solvable} is referred to as 
the set of primary master equations.  
\end{Def}
Then the following proposition holds.
\begin{Proposition}
\label{fact-master-equation-solution}
(Solutions of the primary master equations). 
The solution of \fr{master-equation-solvable} is 
\beq
p(j,t)
=\e^{\,-t}\,p(j,0)+(1-\e^{\,-t})p_{\,\theta}^{\,\eq}(j).
\label{master-equation-solvable-solution}
\eeq
Then it follows that 
\beq
\lim_{t\to\infty}p(j,t)
=p_{\,\theta}^{\,\eq}(j).
\label{asymptotic-limit-probability}
\eeq 
\end{Proposition}
\begin{Proof}
Solving \fr{master-equation-solvable} for $p(j,t)$ with 
a set of given 
initial values $\{p(j,0)\}$,   
one has the explicit form of $p(j,t)$, \fr{master-equation-solvable-solution}. 
From this explicit form, the long-time limit of $p(j,t)$ can  
be given immediately as \fr{asymptotic-limit-probability}.
\qed
\end{Proof}
With this Proposition, one notices the following.
\begin{enumerate}
\item 
Any solution $p$ depends on $\theta$.
\item
Any state transition from $j$ to $j'(\neq j)$ does not occur with this 
set of master equations.  
\item
If $\sum_{j}p(j,0;\theta)=1$, then 
$\sum_{j}p(j,t;\theta)=1$ for any $t>0$. 
\end{enumerate}
Taking into account 1 above,  
$p(j,t)$ is denoted $p(j,t;\theta)$.
\subsection{Dual master equation}
In this subsection, some features of target distribution functions 
$p_{\,\theta}^{\,\eq}$ are discussed first. 
These provide a motivation for introducing 
the dual of the introduced master equations.     
Such master equations and dual master equations 
will be used in Section\,\ref{section-time-development-observables}  
to derive dynamical systems for observables. 
  
Target distribution functions considered in this paper are 
the exponential family, since they are discrete distribution functions.  
This family is a class of probability functions defined as follows. 
Let $\theta=\{\theta^{\,a}\}\in\Theta\subseteq\mbbR^{n}$ be a finite parameter 
set, 
and $\xi$ a set of random variables. 
If a probability function $\check{p}_{\,\theta}$ being parameterized by 
$\theta$ can be written of the form  
\beq
\check{p}_{\,\theta}(\xi)
=\exp(\,\theta^{\,a}\cO_{\,a}(\xi)-\check{\Psi}(\theta)+C(\xi)\,),
\label{exponential-family}
\eeq
with some functions $\{\cO_{\,a}\}, C$ and $\check{\Psi}$, then 
$\check{p}_{\,\theta}$ is said to belong to the exponential family. 
In this paper, the Einstein convention, when an index variable
appears twice in a single term it implies summation of all the values 
of the index, is adapted.
In what follows the function $C$ is assumed to be eliminated.
Also, the following is assumed. 
\begin{Postulate}
\label{postulate-thermodynamic-conjugate-variables-exponential-family}
In \fr{exponential-family},  
$\theta^{\,a}$ and $\mbbE[\cO_{\,a}]$ for each $a$ form a pair of 
dimension-less 
thermodynamic conjugate variables, where $\mbbE[\cO_{\,a}]$ 
is the expectation value of $\cO_{\,a}$ with respect to $\check{p}_{\,\theta}$.
\end{Postulate}
The function $\check{\Psi}$ is used for normalizing $\check{p}_{\,\theta}$, 
and gives the quantities 
\beq
\eta_{\,a}
:=\mbbE[\cO_{\,a}]
=\frac{\partial\check{\Psi}}{\partial \theta^{\,a}},\qquad 
a\in\{1,\ldots,n\}
\label{eta-exponential-family-general}
\eeq
uniquely. Then the correspondence between $\theta^{\,a}$ and $\eta_{\,a}$ 
is one-to-one.  
This uniqueness provides 
a motivation for introducing  
the other expression of 
$p_{\,\theta}^{\,\eq}$. Substituting $\eta=\eta(\theta)$ into 
$p_{\,\theta}^{\,\eq}$, one has the other expression of the target distribution. 
This is denoted by 
$p_{\,\eta}^{\,\eq}$, and the corresponding master equations are  
obtained from 
\fr{master-equation-solvable} as 
\beq
\frac{\partial}{\partial t}p(j,t)
=p_{\,\eta}^{\,\eq}(j)-p(j,t).
\label{master-equation-solvable-dual}
\eeq
This equation is termed in this paper as follows.
\begin{Def}
(Dual master equations).  
The set of equations \fr{master-equation-solvable-dual} 
is referred to as the set of the dual master 
equations associated with \fr{master-equation-solvable}. 
\end{Def}
The primary master equations are used for 
the cases where the set $\theta=\{\theta^{\,a}\}$ 
is kept fixed, and $\eta$ depends on 
random variables.  On the other hand,  
the dual master equations are used for the cases  
where $\eta=\{\eta_{\,a}\}$ in 
\fr{eta-exponential-family-general} is kept fixed, and $\theta$ depends on 
random variables.   

Similar to Proposition\,\ref{fact-master-equation-solution}, one has 
the following.
\begin{Proposition}
\label{fact-master-equation-solution-dual}  
(Solutions of the dual master equations). 
The solution of \fr{master-equation-solvable-dual} is 
\beq
p(j,t)
=\e^{\,-t}\,p(j,0)+(1-\e^{\,-t})p_{\,\eta}^{\,\eq}(j).
\label{master-equation-solvable-solution-dual}
\eeq
Then it follows that 
\beq
\lim_{t\to\infty}p(j,t)
=p_{\,\eta}^{\,\eq}(j).
\label{asymptotic-limit-probability-dual}
\eeq 
\end{Proposition}
\begin{Proof}
A way to prove this is analogous to the proof of 
Proposition\,\ref{fact-master-equation-solution}.
\qed 
\end{Proof}

Accordingly, the solution of this equation is denoted $p(j,t;\eta)$. 

In this paper 
the primary and the dual master equations,  
\fr{master-equation-solvable} and 
\fr{master-equation-solvable-dual}, 
are referred to as 
solvable master equations. 

Before closing this Section, 
attention is concentrated on a spin system 
to show how the general theory developed in this Section  
is applied to a physical model. 
\begin{Example} 
\label{example-kinetic-ising-0}
(Kinetic Ising model without interaction).  
Let $\sigma$ be a spin variable that takes the 
values $\sigma=\pm1$, 
$\scrH$ a constant  magnetic field whose dimension is an energy, 
$T$ the absolute temperature,
and $\theta=\scrH/(k_{\,\B}T)$ with $k_{\,\B}$ being the Boltzmann constant.
Consider the equilibrium system consisting of the one-spin system  
being coupled with a heat bath with $T$.
With these introduced variables, the canonical distribution is given 
by   
\beq
p_{\,\Ising\,\theta}^{\,\eq}(\sigma)
=\exp(\,\theta\,\sigma)/\,Z_{\,\Ising}(\theta),
\label{equilibrium-distribution-Ising-theta}
\eeq
where $Z_{\,\Ising}(\theta)$ is calculated to be 
$$
Z_{\,\Ising}(\theta)
=\sum_{\sigma=\pm1}\exp(\,\theta\,\sigma)
=2\,\cosh(\theta).
$$
When this canonical distribution is chosen to be 
the target distribution function,
the set of primary master equations  
is immediately obtained from \fr{master-equation-solvable} as 
$$
\frac{\partial}{\partial t}\,p_{\,\Ising}(\sigma,t;\theta)
=\exp(\,\theta\,\sigma-\psi_{\,\Ising}^{\,\eq}(\theta))
-p_{\,\Ising}(\sigma,t;\theta),
\qquad 
\sigma
=\pm 1,
$$
where 
\beq
\psi_{\,\Ising}^{\,\eq}(\theta)
:=\ln Z_{\,\Ising}(\theta)
=\ln(2\,\cosh(\theta)).
\label{psi-equilibrium-Ising}
\eeq
Fig.\,\ref{picture-Ising-basics} (a) shows 
the graph of $\psi_{\,\Ising}^{\,\eq}$, and (b) shows a schematic picture of 
the time-development of $p_{\,\Ising}$. Since 
$p_{\,\Ising}(+1,t;\theta)+p_{\,\Ising}(-1,t;\theta)=1$ 
holds for any $t$, 
the dynamics takes place on this one-dimensional line.
\begin{figure}[ht]
\begin{picture}(150,95)
\unitlength 1mm
\put(25,33){(a)}
\put(25,35){\includegraphics[width=3.4cm,angle=-90]{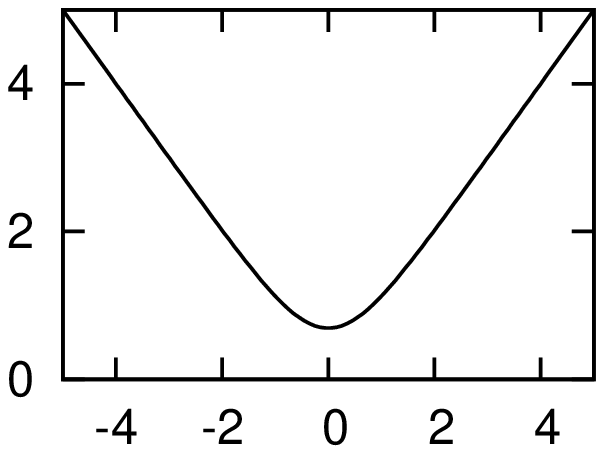}}
\put(45,25){$\psi_{\,\Ising}^{\,\eq}(\theta)$}
\put(50,0){$\theta$}
\put(90,33){(b)}
\put(100,5){\line(1,0){23}}
\put(100,5){\line(0,1){23}}
\put(100,25){\line(1,-1){20}}
\put(100,0){$0$}
\put(120,0){$1$}
\put(97,22){$1$}
\put(125,4){$p_{\,\Ising}(+1,t;\theta)$}
\put(98,30){$p_{\,\Ising}(-1,t;\theta)$}
\put(108,15){\textbullet}
\put(112,18){$(\,p_{\,\Ising\,\theta}^{\,\eq}(+1)\,,p_{\,\Ising\,\theta}^{\,\eq}(-1)\,)$}
\put(101,24){\vector(1,-1){3}}
\put(104,21){\vector(1,-1){3}}
\put(118,7){\vector(-1,1){3}}
\put(115,10){\vector(-1,1){3}}
\end{picture}
\caption{(a) : Graph of $\psi_{\,\Ising}^{\,\eq}$.\qquad\qquad (b): Time-development of $p_{\,\Ising}$.}
\label{picture-Ising-basics}
\end{figure}

The  set of dual master equations  
is derived as follows. First, from 
\fr{eta-exponential-family-general} and \fr{psi-equilibrium-Ising} it follows
that  
$$
\eta
=\tanh\theta.
$$  
Substituting this and the mathematical formula   
$$
\atanh(x)
=\acosh\left(\frac{1}{\sqrt{1-x^2}}\right),\quad |\,x\,|<1,
$$
into $Z_{\,\Ising}(\theta)$, one has 
$$
Z_{\,\Ising}(\theta(\eta))
=2\cosh\left(\atanh\eta\right)
=\frac{2}{\sqrt{1-\eta^{\,2}}},\qquad
|\,\eta\,|<1.
$$
Combining these with \fr{equilibrium-distribution-Ising-theta},
one has 
$$
p_{\,\Ising\,\eta}^{\,\eq}(\sigma)
=\frac{\sqrt{1-\eta^{\,2}}}{2}\,\exp\left(\,\sigma\,\atanh\eta\,\right),
\qquad |\,\eta\,|<1.
$$
Finally, the explicit form of \fr{master-equation-solvable-dual}
is obtained  as  
$$
\frac{\partial}{\partial t}p_{\,\Ising}(\sigma,t;\eta)
=\frac{\sqrt{1-\eta^{\,2}}}{2}\,\exp\left(\,\sigma\,\atanh\eta\,\right)
-p_{\,\Ising}(\sigma,t;\eta),\qquad
\sigma=\pm 1.
$$

The equilibrium free-energy $\cF_{\,\Ising}^{\,\eq}$  is 
written in terms of $\psi_{\,\Ising}^{\,\eq}$ in \fr{psi-equilibrium-Ising} as 
\beq
\cF_{\,\Ising}^{\,\eq}
=-\,k_{\,\B}T\,\psi_{\,\Ising}^{\,\eq},
\label{Ising-free-energy-psi-equilibrium}
\eeq
from which one can interpret $\psi_{\,\Ising}^{\,\eq}$ 
as the negative dimensional-less 
free-energy for this model. 

This system was briefly studied in Ref.\,\cite{Goto2015}, and
is referred to as the kinetic Ising model 
without interaction in this paper.
Note that this system is simplified one 
originally considered in Ref.\cite{Glauber1967}.   

\end{Example}
\section{Time-development of observables}
\label{section-time-development-observables}
In this section differential equations describing 
time-development of 
observables are derived with the solvable master equations 
 under some assumptions. 
Here observable in this paper 
is defined as a function that does not depend on 
a random variable or a state.
Thus expectation values with respect to a probability distribution function  
are observables. 
Then, time-asymptotic limits of such observables are stated.    

\subsection{Expectation values associated with $p(j,t;{\,\theta})$}
In this subsection, the case where 
$\cO_{\,a}$ in \fr{exponential-family} depends on random variables is considered.
First, one defines some expectation values. 
\begin{Def}
(Expectation value with respect to $p(j,t;\theta)$\,).   
Let $\cO_{\,a}:\Gamma\to \mbbR$ be a function 
with $a\in\{1,\ldots,n\}$, and 
$p:\Gamma\times\mbbR\to \mbbR_{\geq 0}$  
a distribution function that 
follows \fr{master-equation-solvable}. Then 
$$
\avg{\cO_{\,a}}_{\,\theta}(t)
:=\sum_{j\in \Gamma}\cO_{\,a}(j)\,p(j,t;\theta),\qquad \mbox{and}\qquad
\avg{\cO_{\,a}}_{\,\theta}^{\,\eq}
:=\sum_{j\in \Gamma}\cO_{\,a}(j)\,p_{\,\theta}^{\,\eq}(j),
$$
are referred to as the expectation value of $\cO_{\,a}$ with respect to $p$,  
and that with respect to $p_{\,\theta}^{\,\eq}$, respectively.
\end{Def}

If a target distribution function belongs to the 
exponential family, then the function $\Psi^{\,\eq}:\Theta\to\mbbR$ with
\beq
\Psi^{\,\eq}(\theta)
:=\ln\left(\,\sum_{j\in \Gamma}\e^{\,\theta^{\,b}\cO_{\,b}(j)}
\,\right),
\label{definition-Psi-eq}
\eeq
plays various roles. Here and in what follows, \fr{definition-Psi-eq} 
is assumed to exist.
In the context of information geometry, this function is referred to as 
the $\theta$-potential. 
Discrete distribution functions 
are considered in this paper 
and it has been known that such distribution functions belong to 
the exponential family, then $\Psi^{\,\eq}$ in 
\fr{definition-Psi-eq} also plays a role throughout this paper.   
The value of the function $\Psi^{\,\eq}(\theta)$ can be interpreted as the 
negative dimension-less free-energy 
(See \fr{Ising-free-energy-psi-equilibrium}).

One then can extend this function to that for nonequilibrium states as 
$\Psi:\Theta\times\mbbR\to\mbbR$ with  
\beq
\Psi(\theta,t)
:=\left(\frac{1}{J^{\,0}}
\sum_{j\in \Gamma}\frac{p(j,t;\theta)}{p_{\theta}^{\,\eq}(j)}\right)
\Psi^{\,\eq}(\theta),\qquad\mbox{where}\qquad
J^{\,0}
:=\sum_{j'\in \Gamma} 1. 
\label{definition-Psi-general-nonequilibrium}
\eeq
Since $p_{\,\theta}^{\,\eq}(j)\neq 0$ and $\Psi^{\,\eq}(\theta)<\infty$  
by assumptions, the function $\Psi$  
\fr{definition-Psi-general-nonequilibrium} exists. 
Extending the idea for the equilibrium case, the function $\Psi$ may be 
interpreted as a nonequilibrium negative dimension-less free-energy. 
Although there could be more suitable 
free-energy for nonequilibrium states,   
\fr{definition-Psi-general-nonequilibrium} is employed in this paper.

One notices the following. 

\begin{Remark}
\label{assumptions-eta-psi}
\begin{enumerate}
\item
Relations among the expectation values of  quantities introduced above 
in the time-asymptotic limit 
are found with \fr{asymptotic-limit-probability} 
as 
$$
\lim_{t\to\infty}\avg{\cO_{\,a}}_{\,\theta}(t)
=\avg{\cO_{\,a}}_{\,\theta}^{\,\eq},
\quad\mbox{and}\quad
\lim_{t\to\infty}\Psi(\theta,t)
=\Psi^{\,\eq}(\theta),
$$
\item
Since a target distribution function belongs to the exponential family,  
it follows that \cite{AN} 
\beq
\avg{\ln p_{\,\theta}^{\,\eq}}_{\,\theta}^{\,\eq}
=\Leg[\Psi^{\,\eq}](\eta),\qquad\mbox{with}\qquad
\eta_{\,a}
=\frac{\partial\Psi^{\,\eq}(\theta)}{\partial \theta^{\,a}}.
\label{eta-derivative-of-Psi}
\eeq
Here $\Leg[\Psi^{\,\eq}]$ is the Legendre transform of $\Psi^{\,\eq}$,  
$$
\Leg[\Psi^{\,\eq}](\eta)
:=\sup_{\theta\in\Theta}\left[\,\theta^{\,a}\,\eta_{\,a}-\Psi^{\,\eq}(\theta)\right].
$$ 
\item
Since a target distribution function is 
a discrete one and thus belongs to the exponential family, 
one has (see \fr{eta-exponential-family-general} and Ref. \cite{AN}) 
\beq
\avg{\cO_{\,a}}_{\,\theta}^{\,\eq}
=\frac{\partial\, \Psi^{\,\eq}(\theta)}{\partial\theta^{\,a}}.
\label{weighted-averages-moment-generating-functions}
\eeq
\end{enumerate}
\end{Remark}
Also, define cross entropy $H$, and negative entropy at equilibrium 
$H^{\,\eq}$ so that   
\beq
H(\theta,t)
:=\sum_{j\in \Gamma}p(j,t;\theta)\ln p_{\,\theta}^{\,\eq}(j)
=\avg{\ln p_{\,\theta}^{\,\eq}}_{\,\theta},
\quad\mbox{and}\quad
H^{\,\eq}(\theta)
:=\sum_{j\in \Gamma}p_{\,\theta}^{\,\eq}(j)\ln p_{\,\theta}^{\,\eq}(j)
=\avg{\ln p_{\,\theta}^{\,\eq}}_{\,\theta}^{\,\eq}.
\label{cross-entropy-theta}
\eeq
From these definitions with \fr{asymptotic-limit-probability},
the time-asymptotic limit of $H$ is obtained as 
$$
\lim_{t\to\infty}H(\theta,t)
=H^{\,\eq}(\theta).
$$

\subsection{Expectation values  associated with $p(j,t;{\,\eta})$}
From discussions on dual master equations, 
it is natural to consider systems where $\theta=\{\theta^{\,a}\}$ depends on 
random variables. 
For this purpose, the set $\theta$ is treated as variables depending on $j$
in this subsection.  

One defines the following.
\begin{Def}
(Expectation value with respect to $p(j,t;\eta)$\,). 
Let $\theta^{\,a}:\Gamma\to\mbbR$ be a function with $a\in\{1,\ldots,n\}$, and 
$p:\Gamma\times\mbbR\to \mbbR_{\geq 0}$  
a distribution function that follows 
\fr{master-equation-solvable-dual}. Then  
$$
\avg{\theta^{\,a}}_{\,\eta}(t)
:=\sum_{j\in \Gamma}\theta^{\,a}(j) p(j,t;\eta),\qquad \mbox{and}\qquad
\avg{\theta^{\,a}}_{\,\eta}^{\,\eq}
:=\sum_{j\in \Gamma}\theta^{\,a}(j) p_{\,\eta}^{\,\eq}(j),
$$
are referred to as the expectation value of $\theta^{\,a}$ with respect to
$p$, and that with respect to $p_{\,\eta}^{\,\eq}$, respectively. 
\end{Def}

One notices the following.
\begin{Remark}
\label{assumptions-eta-phi}
\begin{enumerate}
\item
Relations among the expectation values of quantities introduced above in 
the time-asymptotic limit are found with \fr{asymptotic-limit-probability-dual}
as 
$$
\lim_{t\to\infty}\avg{\theta^{\,a}}_{\,\eta}(t)
=\avg{\theta^{\,a}}_{\,\eta}^{\,\eq}.
$$
\item 
The following relation is satisfied with 
the set of time-independent variables 
$\eta=\{\eta_{\,1},\ldots,\eta_{\,n}\}$ 
\beq
\Leg[\Psi^{\,\eq}](\eta)
=H^{\,\eq}(\theta(\eta)),
\qquad\mbox{with}\qquad
\eta_{\,a}
=\frac{\partial\,\Psi^{\,\eq}(\theta)}{\partial \theta^{\,a}}.
\label{Legendre-transform-Psi-is-H}
\eeq
Here the function $H^{\,\eq}$ as a function of $\eta$ 
is obtained as the following two steps. 
(i) Solving  
$\eta_{\,b}=\partial\Psi^{\,\eq}/\partial\theta^{\,b}$ for $\theta^{\,a}$,
one writes $\theta^{\,a}=\theta^{\,a}(\eta)$. (ii) Substituting 
$\theta^{\,a}=\theta^{\,a}(\eta)$ into $H^{\,\eq}(\theta)$, one has   
$H^{\,\eq}(\theta(\eta))$.
Letting  $\Phi^{\,\eq}$ be the Legendre transform of $\Psi^{\,\eq}$, one 
can write \fr{Legendre-transform-Psi-is-H} as  
$$
\Phi^{\,\eq}(\eta)
=H^{\,\eq}(\theta(\eta)).
$$ 
In the context of information geometry, the function $\Phi^{\,\eq}$ 
is referred to as the $\eta$-potential.
\item
It can be shown that \cite{AN} 
\beq
\avg{\theta^{\,a}}_{\,\theta}^{\,\eq}
=\frac{\partial\,\Phi^{\,\eq}}{\partial \eta_{\,a}},
\label{theta-by-Phi}
\eeq
and 
\beq
\frac{\partial^{\,2}\Psi^{\,\eq}}{\partial \theta^{\,a}\partial \theta^{\,j}}\,
\frac{\partial^{\,2}\Phi^{\,\eq}}{\partial \eta_{\,j}\partial \eta_{\,b}}
=\delta_{\,a}^{\,b},
\label{Psi-Phi-matrix-inverse}
\eeq
where $\delta_{\,a}^{\,b}$ is the Kronecker delta 
giving unity for $a=b$, and zero otherwise. 
\end{enumerate}
\end{Remark}
Similar to \fr{cross-entropy-theta}, one defines
$$
H(\eta,t)
:=\sum_{j\in \Gamma}p(j,t;\eta)\ln p_{\,\eta}^{\,\eq}(j)
=\avg{\ln p_{\,\eta}^{\,\eq}}_{\,\eta},\qquad\mbox{and}\qquad
H^{\,\eq}(\eta)
:=\sum_{j\in \Gamma}p_{\,\eta}^{\,\eq}(j)\ln p_{\,\eta}^{\,\eq}(j)
=\avg{\ln p_{\,\eta}^{\,\eq}}_{\,\eta}^{\,\eq}.
$$
By definition, it follows that $H^{\,\eq}(\theta(\eta))=H^{\,\eq}(\eta)$, where 
$\theta(\eta)$ is obtained by solving 
$\eta^{\,b}=\partial\Psi^{\,\eq}/\partial\theta^{\,b}$ for $\theta^{\,a}$.

Explicit forms of the expectation values 
and ones of Remarks\,\ref{assumptions-eta-psi} and \ref{assumptions-eta-phi} 
for Example in 
Section\,\ref{section-solvable-heat-bath} can be shown as follows.
\begin{Example}
Consider the kinetic Ising model without interaction introduced in 
Example in Section\,\ref{section-solvable-heat-bath}. 
Choose $\cO$ as  
$$
\cO(\sigma)
=\sigma,
$$ 
from which one has the expectation value of $\cO$ with respect to 
$p_{\,\theta}^{\,\eq}$ as 
$$
\avg{\sigma}_{\,\theta}^{\,\eq}
=\sum_{\sigma=\pm1}\sigma\, p_{\,\theta}^{\,\eq}(\sigma)
=\frac{\e^{\,\theta}-\e^{\,-\,\theta}}{Z_{\,\Ising}(\theta)}
=\tanh(\theta).
$$
The $\theta$-potential $\Psi^{\,\eq}$ defined in \fr{definition-Psi-eq}  is 
calculated to be 
$$
\Psi_{\,\Ising}^{\,\eq}(\theta)
=\ln\left(\sum_{\sigma=\pm 1}\e^{\,\theta\,\sigma}\right)
=\ln\left(\,2\cosh(\theta)\,\right)
=\psi_{\,\Ising}^{\,\eq}(\theta),
$$
where $\psi_{\,\Ising}^{\,\eq}(\theta)$ has been defined in 
\fr{psi-equilibrium-Ising}.
Then,  
one has $\eta$ in \fr{eta-derivative-of-Psi} as 
\beq
\eta
=\frac{\dr\Psi_{\,\Ising}^{\,\eq}(\theta)}{\dr\theta}
=\tanh(\theta).
\label{Ising-eta-theta}
\eeq
Also it turns out that $\Psi_{\,\Ising}^{\,\eq}$ is strictly convex due to 
\beq
\frac{\dr^{\,2}\Psi_{\,\Ising}^{\,\eq}(\theta)}{\dr\theta^{\,2}}
=\frac{\dr\eta}{\dr\theta}
=\sech^{\,2}(\theta)>0.
\label{Psi-Ising-equilibrium-convex}
\eeq
The negative entropy at equilibrium $H_{\,\Ising}^{\,\eq}$ is calculated to be 
\beq
H_{\,\Ising}^{\,\eq}(\theta)
=\sum_{\sigma=\pm 1}p_{\,\theta}^{\,\eq}(\sigma)\ln p_{\,\theta}^{\,\eq}(\sigma)
=\sum_{\sigma=\pm 1}p_{\,\theta}^{\,\eq}(\sigma)\left(\,\theta\sigma
-\psi_{\,\Ising}(\theta)\,\right)
=\theta\tanh(\theta)-\ln\left(\,2\,\cosh(\theta)\,\right).
\label{Ising-H-equilibrium}
\eeq
The relation \fr{Legendre-transform-Psi-is-H} is verified for this Example 
below.
First, the Legendre transform $\Leg\left[\Psi_{\,\Ising}^{\,\eq}\right](\eta)$ 
is calculated as   
$$
\Leg\left[\Psi_{\,\Ising}^{\,\eq}\right](\eta)
=\sup_{\theta}\left[\theta\,\eta-\Psi_{\,\Ising}^{\,\eq}(\theta)\right]
=\left[\theta\,\eta-\Psi_{\,\Ising}^{\,\eq}(\theta)\right]_{\,\theta=\atanh\eta}
=\eta\,\atanh\eta-\ln\left[\,2\cosh(\atanh\eta)\right].
$$
Second,  
it follows from \fr{Ising-eta-theta} and \fr{Ising-H-equilibrium} that 
$\Leg\left[\Psi_{\,\Ising}^{\,\eq}\right](\eta)=H_{\,\Ising}^{\,\eq}(\theta(\eta))$. 
They are denoted by $\Phi_{\,\Ising}^{\,\eq}(\eta)=\Leg\left[\Psi_{\,\Ising}^{\,\eq}\right](\eta)=H_{\,\Ising}^{\,\eq}(\theta(\eta))$. 
Another expression of $\Phi_{\,\Ising}^{\,\eq}(\eta)$ is obtained from  
$$
\atanh(x)
=\frac{1}{2}\ln\left(\frac{1+x}{1-x}\right),\qquad\mbox{and}\qquad
\atanh(x)
=\acosh\left(\frac{1}{\sqrt{1-x^2}}\right),\quad |\,x\,|<1,
$$
as 
$$
\Phi_{\,\Ising}^{\,\eq}(\eta)
=\Leg\left[\Psi_{\,\Ising}^{\,\eq}\right](\eta)
=\frac{\eta}{2}\ln\left(\frac{1+\eta}{1-\eta}\right)
+\frac{1}{2}\ln(1-\eta^2)-\ln 2,
\qquad |\,\eta\,|<1, 
$$
where $-1<\eta<1$ coming from 
\fr{Ising-eta-theta} has been applied. 
Similar to $\Psi_{\,\Ising}^{\,\eq}$, the function $\Phi_{\,\Ising}^{\,\eq}$ 
is shown to be strictly convex directly. It follows from 
$$
\frac{\dr H_{\,\Ising}^{\,\eq}}{\dr \theta}
=\theta\,\sech^2\theta,\qquad\mbox{and}\qquad
\frac{\dr H_{\,\Ising}^{\,\eq}}{\dr \eta}
=\frac{\dr H_{\,\Ising}^{\,\eq}}{\dr \theta}\frac{\dr\theta}{\dr\eta}
=\theta(\eta),
$$
that
\beq
\frac{\dr^{\,2} \Phi_{\,\Ising}^{\,\eq}}{\dr \eta^{\,2}}
=\frac{\dr\theta}{\dr\eta}
=\cosh^2\left[\,\theta(\eta)\,\right]
=\cosh^2\left[\,\atanh(\eta)\,\right]
=\frac{1}{1-\eta^{\,2}}
>0.
\label{Phi-Ising-equilibrium-convex}
\eeq

The relation \fr{Psi-Phi-matrix-inverse} for this Example is verified by  
combining \fr{Psi-Ising-equilibrium-convex}  and 
\fr{Phi-Ising-equilibrium-convex} as 
$$
\frac{\dr^2 \Psi_{\,\Ising}^{\,\eq}}{\dr \theta^2}
\frac{\dr^2 \Phi_{\,\Ising}^{\,\eq}}{\dr \eta^2}
=1.
$$
\end{Example} 

Before closing this subsection, physical interpretations of $\theta^{\,a}$ and 
$\avg{\theta^{\,a}}_{\,\eta}$ are argued by taking a simple example.
Identify $\theta=\scrH/(k_{\,\B} T)$,  
where $\scrH$ is a fixed constant magnetic field and 
$T$ a temperature of a fixed heat bath.
Then, $\theta$ is roughly speaking the inverse temperature of the heat bath.
It should be emphasized that $T$ does not depend on time, and thus it is 
constant. 
On the other hand, one can introduce another 
temperature $\avg{T^{\,\prime}}^{\,\prime}$ that is proportional to  
the expectation value of the squared velocity of 
a particle $\avg{v^{\,2}}^{\,\prime}$, 
where $\avg{\cdots}^{\,\prime}$ has been introduced appropriately.  
In a gas case, 
such a relation is 
$\avg{T^{\,\prime}}^{\,\prime}=m\avg{v^{\,2}}^{\,\prime}/(3\,k_{\,\B})$, 
where $m$ is the mass of a particle.
This leads to 
$\avg{\theta}^{\,\prime}=\scrH/(\,k_{\,\B}\avg{T^{\,\prime}}^{\,\prime}\,)$,  
which is not constant in time in general.
Although $\avg{\cdots}^{\,\prime}$ is not same as $\avg{\cdots}_{\,\eta}$, 
one can generalize this example to the case where 
 $\{\theta^{\,a}\}$ is a fixed parameter set and  
$\{\avg{\theta^{\,a}}_{\,\eta}\}$ an averaged one.

\subsection{Closed dynamical systems}
As shown below a set of differential equations 
for $\{\avg{\cO_{\,a}}_{\,\theta}\}$ and 
$\Psi$ can be found in a closed form.
When expectation values depending on time are seen as 
variables that satisfy differential equations, they are referred to as 
expectation variables in this paper.
\begin{Proposition}
\label{moment-dynamics}
(Dynamical system obtained from the primary master equations).    
Let $\theta$ be a time-independent parameter set characterizing a 
discrete target distribution function $p_{\,\theta}^{\,\eq}$.
Then $\{\avg{\cO_{\,a}}_{\,\theta}\}$ and $\Psi$ are solutions to 
the differential equations on $\mbbR^{\,2n+1}$
$$
\frac{\dr}{\dr t}\theta^{\,a}
=0,\qquad
\frac{\dr}{\dr t}
\avg{\cO_{\,a}}_{\,\theta}
=-\,\avg{\cO_{\,a}}_{\,\theta}
+\frac{\partial\,\Psi^{\,\eq}}{\partial\theta^{\,a}},\quad\mbox{and}\quad 
\frac{\dr}{\dr t}\Psi
=-\,\Psi+\Psi^{\,\eq}.
$$
\end{Proposition}
\begin{Proof}
Non-trivial parts of this proof is for the second and third differential  
equations.
A proof that the equation for $\avg{\cO_{\,a}}_{\,\theta}$  
holds is given first.  
It follows from \fr{master-equation-solvable} 
and \fr{weighted-averages-moment-generating-functions}
that  
$$
\frac{\dr}{\dr t}\avg{\cO_{\,a}}_{\,\theta}
=\sum_{j\in \Gamma}\cO_{\,a}(j)\,
\frac{\partial p(j,t;\theta)}{\partial t}
=\sum_{j\in \Gamma}\cO_{\,a}(j)
\left[\,p_{\,\theta}^{\,\eq}(j)-p(j,t;\theta)
\,\right]
=-\,\avg{\cO_{\,a}}_{\,\theta}
+\frac{\partial\,\Psi^{\,\eq}}{\partial\theta^{\,a}}.
$$
Then, a proof that the equation for $\Psi$ holds is given below. 
It follows from \fr{master-equation-solvable}
and \fr{definition-Psi-general-nonequilibrium} that 
$$
\frac{\dr}{\dr t}\Psi
=\left(\frac{1}{J^{\,0}}\sum_{j'\in \Gamma}\frac{1}{p_{\theta}^{\,\eq}(j')}
\frac{\partial\,p(j',t;\theta)}{\partial t}
\right)
\Psi^{\,\eq}
=\left(\frac{1}{J^{\,0}}
\sum_{j'\in \Gamma}\frac{p_{\theta}^{\,\eq}(j')-p(j',t;\theta)\,}{p_{\,\theta}^{\,\eq}(j')}
\right)
\Psi^{\,\eq}
=-\,\Psi+\Psi^{\,\eq}.
$$
\qed
\end{Proof}
\begin{Remark}
The explicit time-dependence for this system is obtained as 
$\theta^{\,a}(t)=\theta^{\,a}(0)$, 
$$
\avg{\cO_{\,a}}_{\,\theta}(t)
=\e^{\,-\,t}\left[\,\avg{\cO_{\,a}}_{\,\theta}(0)
-\frac{\partial\Psi^{\,\eq}}{\partial \theta^{\,a}}\,\right]
+\frac{\partial\Psi^{\,\eq}}{\partial \theta^{\,a}},
\qquad\mbox{and}\qquad
\Psi(\theta,t)
=\e^{\,-\,t}\left[\,\Psi(0)-\Psi^{\,\eq}(\theta)\,\right]
+\Psi^{\,\eq}(\theta).
$$
From these and \fr{weighted-averages-moment-generating-functions},  
one can also verify that the time-asymptotic limit 
of these variables are those defined at equilibrium :
$$
\lim_{t\to\infty}\avg{\cO_{\,a}}_{\,\theta}\,(t)
=\avg{\cO_{\,a}}_{\,\theta}^{\,\eq},\quad\mbox{and}\quad
\lim_{t\to\infty}\Psi(\theta,t)
=\Psi^{\,\eq}(\theta).
$$
\end{Remark}

For the later convenience, this dynamical system is termed follows.
\begin{Def}
(Primary moment dynamical system).  
The dynamical system in Proposition\,\ref{moment-dynamics}
 is referred to as a (primary) moment dynamical system.
\end{Def}

Similar to the primary   
moment dynamical system, one has the following dynamical system.
\begin{Proposition}
\label{parameter-dynamics}
(Dynamical system obtained from the dual master equations).   
Let $\{\eta_{\,a}\}$ be a set of time-independent variables 
satisfying the second equation of 
\fr{eta-derivative-of-Psi}. Then 
$\avg{\theta^{\,a}}_{\,\eta}(t)$ and $H(\eta,t)$ are 
solutions to the differential equations on $\mbbR^{\,2n+1}$
$$
\frac{\dr}{\dr t}\avg{\theta^{\,a}}_{\,\eta}
=-\,\avg{\theta^{\,a}}_{\,\eta}
+\frac{\partial\, \Phi^{\,\eq}}{\partial \eta_{\,a}},\qquad
\frac{\dr}{\dr t}\eta_{\,a}
=0,\qquad\mbox{and}\qquad
\frac{\dr}{\dr t}H
=-\,H+\Phi^{\,\eq}(\eta).
$$
\end{Proposition}
\begin{Proof}
Non-trivial parts of this proof is for the first and third differential  
equations.
A proof that the equation for
 $\avg{\theta^{\,a}}_{\,\eta}$ holds is given first.
It follows from \fr{master-equation-solvable-dual} and 
\fr{theta-by-Phi} that 
$$
\frac{\dr}{\dr t}\avg{\theta^{\,a}}_{\,\eta}
=\sum_{j\in \Gamma}\theta^{\,a}(j)
\frac{\partial\, p(j,t;\eta)}{\partial t}
=\sum_{j\in \Gamma}\theta^{\,a}(j)\left[\,
p_{\,\eta}^{\,\eq}(j)-p(j,t;\eta)
\,\right]
=-\,\avg{\theta^{\,a}}_{\,\eta}
+\frac{\partial\, \Phi^{\,\eq}}{\partial \eta_{\,a}}.
$$
Then, a proof that the equation for $H(\eta,t)$ holds is given as
$$
\frac{\dr H}{\dr t}
=\sum_{j\in \Gamma}
\frac{\partial\, p(j,t;\eta)}{\partial t}\ln 
 p_{\,\eta}^{\,\eq}(j)
=\sum_{j\in \Gamma}\left[\,
p_{\,\eta}^{\,\eq}(j)-p(j,t;\eta)
\,\right]
\ln p_{\,\eta}^{\,\eq}(j)
=-\,H+\Phi^{\,\eq}(\eta).
$$
\qed
\end{Proof}
\begin{Remark}
The explicit time-dependence for this system is obtained as 
$\eta_{\,a}(t)=\eta_{\,a}(0)$, 
$$
\avg{\theta^{\,a}}_{\,\eta}(t)
=\e^{\,-\,t}\left[\,\avg{\theta^{\,a}}_{\,\eta}(0)
-\frac{\partial\,\Phi^{\,\eq}}{\partial\, \eta_{\,a}}\,\right]
+\frac{\partial\,\Phi^{\,\eq}}{\partial \eta_{\,a}},
\qquad\mbox{and}\qquad
H(\eta, t)
=\e^{\,-\,t}\left[\,H(0)-\Phi^{\,\eq}(\eta)\,\right]
+\Phi^{\,\eq}(\eta).
$$
From these and \fr{theta-by-Phi}, 
 one can also verify that the time-asymptotic limit 
of these variables  
are those defined at equilibrium  :
$$
\lim_{t\to\infty}\avg{\theta^{\,a}}_{\,\eta}(t)
=\avg{\theta^{\,a}}_{\,\eta}^{\,\eq},\quad\mbox{and}\quad
\lim_{t\to\infty}H(\eta,t)
=\Phi^{\,\eq}(\eta).
$$
\end{Remark}

For the later convenience, this dynamical system is termed.
\begin{Def}
(Dual moment dynamical system).  
The dynamical system in Proposition\,\ref{parameter-dynamics}
 is referred to as a dual moment dynamical system.
\end{Def}

In the following, it is shown 
how the general theory developed above is applied by focusing on the system 
stated in  Example in Section\,\ref{section-solvable-heat-bath}. 

\begin{Example}
Consider the kinetic Ising model without interaction introduced in 
Example in Section\,\ref{section-solvable-heat-bath}. 
Choose $\cO$ to be $\cO(\sigma)=\sigma,$ 
from which one has the expectation value of 
$\sigma$ with respect to $p_{\,\theta}^{\,\eq}$ as 
$\avg{\sigma}_{\,\theta}^{\,\eq}=\tanh(\theta)$.
The primary  moment dynamical system   
discussed in Proposition\,\ref{moment-dynamics} 
for this case is 
$$
\frac{\dr \theta}{\dr t}
=0,\qquad
\frac{\dr \avg{\sigma}_{\,\theta}}{\dr t}
=-\,\avg{\sigma}_{\,\theta}+\tanh(\theta),\qquad\mbox{and}\qquad 
\frac{\dr \Psi_{\,\Ising}}{\dr t}
=-\,\Psi_{\,\Ising}+\ln\left(\,2\,\cosh(\theta)\,\right).
$$
The dual moment dynamical system discussed 
in Proposition\,\ref{parameter-dynamics}
for this case is 
$$
\frac{\dr \avg{\theta}_{\,\eta}}{\dr t}
=-\,\avg{\theta}_{\,\eta}+\atanh\eta,\qquad
\frac{\dr \eta}{\dr t}
=0,\quad\mbox{and}\quad 
\frac{\dr H_{\,\Ising}}{\dr t}
=-\,H_{\,\Ising}+
\eta\,\atanh\eta-\ln[\,2\,\cosh(\atanh\eta)\,].
$$
Since $\theta$ is chosen as $\theta=\scrH/(k_{\,\B}T)$ for this Example, 
the equations above correspond to the case where temperature depends on time, 
$t$,   if $\scrH$ is constant.  
\end{Example}
\section{Geometric description of dynamical systems }
\label{section-geometry-solvable-extended-heat-bath}
For the model used in this paper,  
the equilibrium states can be geometrized with information geometry, since 
the target distribution functions belong to the exponential family \cite{AN}.  
Several geometrization of nonequilibrium states for some 
models and methods  
have been proposed \cite{Takabatake2004,Nagaoka2005,Nielsen2018}.  
Yet, suffice to say that there
remains no general consensus on how best to extend information geometry of 
equilibrium states to a geometry expressing nonequilibrium states.
In this section, a geometrization of 
nonequilibrium states is proposed for the solvable 
master equations.  

Also it is shown that such a geometry 
is consistent with information geometry in the sense that the time-asymptotic 
limit of solutions to the primary and dual  
master equations is described in information geometry.
Here one notices that manifolds and geometric framework expressing 
equilibrium states should be obtained in the limiting case of 
nonequilibrium ones. Thus, 
the geometry for expressing nonequilibrium states 
should be wider than the geometry expressing equilibrium states 
in some sense. 
One such a wider geometry is contact geometry \cite{Goto2015,Bravetti2015}.

\subsection{Geometry of equilibrium states}
\label{section-geometry-equilibrium-states}
In what follows the relation between contact geometry and information geometry
is briefly reviewed.

Let $\cC$ be a $(2n+1)$-dimensional manifold, ($n=1,2,\ldots$). 
If a one-form  $\lambda$ on $\cC$ is provided and satisfies 
\beq
\lambda\wedge\underbrace{\dr\lambda\wedge\cdots\wedge\dr\lambda}_{n}\neq 0,
\label{contact-manifold-volume-form}
\eeq 
then the pair $(\cC,\lambda)$ is referred to as a contact manifold, 
and $\lambda$ a contact one-form. 
It has been known that there exists a special set of coordinates 
$(x,y,z)$ with $x=\{x^{\,1},\ldots,x^{\,n}\}$ and $y=\{y_{\,1},\ldots,y_{\,n}\}$  
such that $\lambda=\dr z-y_{\,a}\dr x^{\,a}$.
The existence of  
such coordinates is guaranteed mathematically stated as the 
Darboux theorem \cite{Silva2008}.  
The Legendre submanifold $\cA\subset\cC$ 
is an $n$-dimensional manifold where $\lambda|_{\,\cA}=0$ holds.
One can verify that 
\beq
\cA_{\,\varpi}
=\left\{\,(x,y,z)\ \bigg|\ 
y_{\,a}
=\frac{\partial\varpi}{\partial x^{\,a}},\quad
\mbox{and}\quad 
z=\varpi(x)\,\right\}, 
\label{Legendre-submanifold-psi}
\eeq
is a Legendre submanifold, where $\varpi:\cC\to\mbbR$  is a function of $x$ 
on $\cC$. The submanifold $\cA_{\,\varpi}$ is referred to as the 
Legendre submanifold generated by $\varpi$, and is used for describing 
equilibrium thermodynamic systems \cite{Mrugala2000}. It should be noted that 
how statistical mechanics dealing with distribution functions 
adopts Legendre submanifolds 
was investigated \cite{Mrugala1990,Bravetti2015}.

Equilibrium states, or equivalently target states, 
 are identified with the Legendre submanifolds generated by functions    
 in the context of 
geometric thermodynamics \cite{Mrugala1978,Mrugala2000}.  
Besides, in the context of information geometry, equilibrium states are 
identified with dually flat spaces.  
Combining these identifications, one has the following. 
\begin{Proposition}
(A contact manifold and a strictly convex function induce a dually flat 
space, \cite{Goto2015}).  
Let $(\cC,\lambda)$ be a contact manifold, $(x,y,z)$ a set of coordinates such 
that $\lambda=\dr z-y_{\,a}\dr x^{\,a}$ with $x=\{x^{\,1},\ldots,x^{\,n}\}$ and 
$y=\{y_{\,1},\ldots,y_{\,n}\}$, 
and $\varpi$ a strictly convex function depending only on $x$.
If the Legendre submanifold generated by $\varpi$ is simply connected, 
then $((\cC,\lambda),\varpi)$ induces the $n$-dimensional dually flat space
\end{Proposition}

Dually flat space is defined in information geometry, and 
this space consists of 
a manifold $\cM$, a (pseudo-) Riemannian metric tensor field $g^{\,\F}$, 
and flat connections $\nabla,\nabla^{\,*}$ that satisfy 
\beq
X\left[\,g^{\,\F}(Y,Z)\,\right]
=g^{\,\F}(\nabla_XY,Z)+g^{\,\F}(Y,\nabla_X^{\,*}Z),\qquad 
\forall\,X,Y,Z\in T\cM.
\label{dually-connection-definition}
\eeq 
Thus  a  dually flat space is a 
quadruplet $(\cM,g^{\,\F},\nabla,\nabla^{\,*})$. 

To apply this Proposition to physical systems, 
the coordinate sets $x$ and $y$ are chosen such that  $x^{\,a}$ and $y_{\,a}$ 
form a thermodynamic conjugate pair for each $a$. 
Here it is assumed that 
such thermodynamic variables can be defined even for nonequilibrium states, 
and they are consistent with those variables defined at equilibrium. 
In addition to this, 
the physical dimension of $\varpi$ should be equal to that of 
$y_{\,a}\,\dr x^{\,a}$. 

How to specify a function $\varpi$ in Proposition is given as follows. 
In mathematical statistics, given a probability 
distribution function 
$\check{p}_{\,\theta}$ parameterized by $\theta$,  
the Fisher information matrix is defined. 
Each component is defined by   
\beq
g_{\,ab}^{\,\F}(\theta)
=\mbbE\left[\,\frac{\partial \ln \check{p}_{\,\theta}}{\partial \theta^{\,a}}
\frac{\partial \ln \check{p}_{\,\theta}}{\partial \theta^{\,b}}\,\right].
\label{Fisher-matrix-component-with-p}
\eeq
For the exponential family, 
these  matrix components 
can  be obtained not only by calculating \fr{Fisher-matrix-component-with-p},  
but also by differentiating  $\Psi^{\,\eq}$ in \fr{definition-Psi-eq}  
with respect to $\theta$ twice \cite{AN}. This $\Psi^{\,\eq}$ and 
its Legendre transform $\Phi^{\,\eq}=\Leg[\Psi^{\,\eq}]$, 
a $\theta$-potential and an $\eta$-potential, are chosen as 
$\varpi$ in this Proposition.
This choice is a part of  procedure linking the space of distribution functions 
and a contact manifold.

Instead of discussing physical meaning of the Proposition, 
we explain how this Proposition applies to a physical system 
by analyzing Example in Section\,\ref{section-solvable-heat-bath}. 
\begin{Example}
Consider the kinetic Ising model without interaction introduced in 
Example in Section\,\ref{section-solvable-heat-bath}. 
Let $\cC$ be the $3$-dimensional manifold
with $\cC\cong \mbbR\times (-1,1)\times\mbbR$, 
and $\lambda=\dr z-y\,\dr x$.
Identify the coordinates such that  $x=\theta$, 
$y=\avg{\sigma}_{\,\theta}$,  and $z=\Psi$, 
then the condition \fr{contact-manifold-volume-form} is 
$\dr\theta\wedge\dr\avg{\sigma}_{\,\theta}\wedge\dr\Psi\neq 0$. 
Then the Legendre submanifold  generated by 
$\Psi_{\,\Ising}^{\,\eq}$,
\beq
\cA_{\,\Psi_{\,\Ising}^{\,\eq}}
=\left\{\,(\,\theta,\avg{\sigma}_{\,\theta},\Psi\,)\,
\bigg|\,\avg{\sigma}_{\,\theta}=\frac{\dr\Psi_{\,\Ising}^{\,\eq}}{\dr\theta} 
\quad\mbox{and}\quad \Psi=\Psi_{\,\Ising}^{\,\eq}(\theta)
\,\right\},
\label{Ising-Legendre-submanifold-Psi}
\eeq
is a suitable submanifold expressing   
the equilibrium state. It is known that 
equilibrium thermodynamic
systems can be described in terms of information geometry.
For example, the Fisher metric tensor field 
and the expectation coordinate  on $\cA_{\,\Psi_{\,\Ising}^{\,\eq}}$ 
are 
$$
g_{\,\Ising}^{\,\F\,\Psi}
=g_{\,\Ising,\theta\theta}^{\,\F}\,
\dr\theta\otimes\dr\theta
=\sech^{\,2}\theta\,\dr\theta\otimes\dr\theta,
\qquad\mbox{and}\qquad
\eta=\frac{\dr\,\Psi_{\,\Ising}^{\,\eq}}{\dr\theta}
=\tanh(\theta),
$$
respectively, where  \fr{Psi-Ising-equilibrium-convex}, 
\fr{Ising-eta-theta},  and  
$$
g_{\,\Ising,\theta\theta}^{\,\F}
=\frac{\dr^{\,2}{\Psi_{\,\Ising}^{\,\eq}}}{\dr\theta^{\,2}}, 
$$
have been used. 
Combining these and the first equality of 
\fr{Phi-Ising-equilibrium-convex}, 
one verifies that $\theta$ and $\eta$ are dual with respect to 
$g_{\,\Ising}^{\,\F}$ in the sense that \cite{AN} 
$$
g_{\,\Ising}^{\,\F}\left(\frac{\partial}{\partial \theta},
\frac{\partial}{\partial \eta}\right)
=\left(\,\sech^2\theta\,\right)\,\frac{\dr\theta}{\dr\eta}
=1.
$$
Since the Fisher metric tensor filed is a type of 
Riemannian metric tensor field 
and it is known that a Riemannian metric tensor field induces the Levi-Civita 
connection uniquely \cite{Nakahara1990},    
$g_{\,\Ising}^{\,\F}$ induces the Levi-Civita 
connection $\nabla^{(0)}$. Its connection component 
$\Gamma_{\,\theta\theta}^{\,(0)\,  \theta}$ is  such that   
$\nabla_{\partial}^{\,(0)}\partial=\Gamma_{\,\theta\theta}^{\,(0)\, \theta}\partial$,  
($\partial:=\partial/\partial\theta$), and its explicit form is calculated 
as \cite{Nakahara1990}  
$$
\Gamma_{\,\theta\theta}^{\,(0)\,  \theta}
=\frac{1}{2}\,g_{\,\Ising}^{\,\F\,\theta\theta}\frac{\dr}{\dr\theta}\,
g_{\,\Ising,\theta\theta}^{\,\F}
=-\,\tanh\theta,\qquad (\mbox{no sum}),
$$  
where 
$g_{\,\Ising}^{\,\F\,\theta\theta}=(g_{\,\Ising,\theta\theta}^{\,\F})^{\,-1}=\cosh^{\,2}\theta$. 
Then one can show 
$$
\Gamma_{\,\theta\theta\theta}^{\,(0)}
:=g_{\,\Ising,\theta\theta}^{\,\F}\,\Gamma_{\,\theta\theta}^{\,(0)\, \theta}
=\frac{1}{2}C_{\,\Ising,0}^{\,\F}
=-\,\sech^{\,2}\theta\tanh\theta,\qquad (\mbox{no sum})
$$ 
where $C_{\,\Ising,0}^{\,\F}$ is the component of the cubic-form, and 
is  related to the Levi-Civita connection.  
The cubic-form is defined in information geometry 
as $C^{\,\F}:=\nabla\,g^{\,\F}$   with 
$\nabla$ being a connection \cite{MatsuzoeHenmi2013}. 
If $\nabla$ is flat, then its component expression for 
this Example is obtained as  
$$
C_{\,\Ising}^{\,\F}
=C_{\,\Ising,0}^{\,\F}\,\dr\theta\otimes \dr\theta\otimes\dr\theta,\qquad
\mbox{where}\qquad
C_{\,\Ising,0}^{\,\F}
=\frac{\dr^{\,3}\,\Psi_{\,\Ising}^{\,\eq}}{\dr\theta^{\,3}}
=-\,2\,\sech^{\,2}\theta\tanh\theta.
$$ 
The $\alpha$-connection defined in information geometry gives dual connections.
Since the target distribution function belongs to the exponential family, 
one can show that the component of the $\alpha$-connection 
$\Gamma_{\,\theta\theta}^{\,(\alpha)\, \theta}$ 
can be obtained from 
$\Gamma_{\,\theta\theta}^{\,(\alpha)\, \theta}=g_{\,\Ising}^{\,\F\,\theta\theta}\,\Gamma_{\,\theta\theta\theta}^{\,(\alpha)}$ and 
$$
\Gamma_{\,\theta\theta\theta}^{\,(\alpha)}
=\frac{1-\alpha}{2}C_{\,\Ising,0}^{\,\F},\qquad \alpha\in\mbbR.
$$ 
From this, it follows that $\nabla^{\,(1)}$ is flat with $\theta$ being 
its coordinate.  
Combining the equations above, one has the component expression of 
\fr{dually-connection-definition},  
$$
\frac{\dr}{\dr\theta}\,g_{\,\Ising,\theta\theta}^{\,\F}
=\Gamma_{\,\theta\theta\theta}^{\,(\alpha)}+\Gamma_{\,\theta\theta\theta}^{\,(-\alpha)}.
$$
Furthermore,
it can be shown that $\nabla^{(-1)}$ is flat with $\eta$ being its 
coordinate \cite{AN}. 
To summarize, the quadruplet 
$(\cA_{\,\Psi_{\,\Ising}^{\,\eq}},g_{\,\Ising}^{\,\F},\nabla^{\,(1)},\nabla^{\,(-1)})$
is a dually flat space, and is induced from the contact manifold 
$\cC$, $\lambda$, 
and $\Psi_{\,\Ising}^{\,\eq}$.

Instead of the identification employed above, 
one can identify the coordinates such that  
$x=\eta, y=\avg{\theta}_{\,\eta}$, and $z=\Phi$ for  
$\cC\cong(-1,1)\times\mbbR^{\,2}$.  
Then, the condition \fr{contact-manifold-volume-form} is 
$\dr\eta\wedge\dr\avg{\theta}_{\,\eta}\wedge\dr\Phi\neq 0$.
In this case the Legendre submanifold generated by $\Phi_{\,\Ising}^{\,\eq}$,  
$$
\cA_{\,\Phi_{\,\Ising}^{\,\eq}}
=\left\{\,(\eta,\avg{\theta}_{\,\eta},\Phi)
\bigg|\,\avg{\theta}_{\,\eta}=\frac{\dr\Phi_{\,\Ising}^{\,\eq}}{\dr\eta},\quad\mbox{and}\quad
\Phi=\Phi_{\,\Ising}^{\,\eq}
\,\right\},
$$
is a suitable submanifold expressing the equilibrium state.
On this submanifold, geometric objects such as the Fisher metric tensor field,
the cubic form, and the $\alpha$-connection, can explicitly be constructed as 
well. For example, the Fisher metric tensor field is 
$$
g_{\,\Ising}^{\,\F\,\Phi}
=g_{\,\Ising}^{\,\F\,\eta\eta}\,
\dr\eta\otimes\dr\eta
=\frac{1}{1-\eta^{\,2}}\,\dr\eta\otimes\dr\eta,\qquad
g_{\,\Ising}^{\,\F\,\eta\eta}
=\frac{\dr^{\,2}{\Phi_{\,\Ising}^{\,\eq}}}{\dr\eta^{\,2}}. 
$$

The relation between 
$\cA_{\,\Psi_{\,\Ising}^{\,\eq}}$ and $\cA_{\,\Phi_{\,\Ising}^{\,\eq}}$ can be argued as 
follows. 
It follows from \fr{Psi-Ising-equilibrium-convex} and 
\fr{Phi-Ising-equilibrium-convex} 
that there exists a diffeomorphism between these 
submanifolds \cite{Goto2015,Bravetti2015}. Let 
$\phi_{\,\Psi\Phi}:\cA_{\,\Psi_{\,\Ising}^{\,\eq}}\to\cA_{\,\Phi_{\,\Ising}^{\,\eq}}$ be such a 
diffeomorphism. Then it follows from straightforward 
calculations \cite{Nakahara1990}
that the pull-back of $g_{\,\Ising}^{\,\F\,\Phi}$ is 
shown to equal to $g_{\,\Ising}^{\,\F\,\Psi}$,  
$$
\phi_{\,\Psi\Phi}^{\,*}\,(\,g_{\,\Ising}^{\,\F\,\Phi}\,)
=g_{\,\Ising}^{\,\F\,\Psi}.
$$

\end{Example}
So far geometry of equilibrium states have been discussed. 
One remaining issue is 
how to give the physical meaning of 
the set outside $\cA_{\,\varpi}$, 
$\cC\setminus\cA_{\,\varpi}$. 
A natural interpretation 
of $\cC\setminus\cA_{\,\varpi}$ 
would be some set of nonequilibrium states, and 
is discussed in the next subsection.

\subsection{Geometry of nonequilibrium states}
As shown in Proposition\ref{fact-master-equation-solution}, 
initial states approach to the equilibrium state as time develops.
This can be reformulated on contact manifolds.  
In the contact geometric framework of nonequilibrium thermodynamics, 
the equilibrium state is identified with a Legendre submanifold. 
Then, as it has been clarified in Refs. \cite{Goto2015} and 
\cite{Bravetti2015},  
this asymptotic behavior can be identified with a class of 
contact Hamiltonian vector fields on a contact manifold. 
Here, contact Hamiltonian vector fields are analogous to symplectic 
vector fields on symplectic manifolds, 
and symplectic vector fields correspond to canonical 
equations of motion in Hamiltonian mechanics. 
This statement on a class of contact Hamiltonian vector fields 
can be summarized as follows.
\begin{Proposition}
\label{proposition-goto-2015}
(Legendre submanifold as an attractor, \cite{Goto2015}). 
Let $(\cC,\lambda)$ be a $(2n+1)$-dimensional contact manifold with $\lambda$ 
being a contact form, $(x,y,z)$ its coordinates so that 
$\lambda=\dr z-y_{\,a}\dr x^{\,a}$ with 
$x=\{x^{\,1},\ldots,x^{\,n}\}$, $y=\{y_{\,1},\ldots,y_{\,n}\}$, and 
$\varpi$ a function
depending only on $x$. Then, one has 
\begin{enumerate}
\item
The contact Hamiltonian vector field associated with 
the contact Hamiltonian $h:\cC\to\mbbR$ such that
\beq
h(x,y,z)
=\varpi(x)-z,
\label{contact-Hamiltonian-for-relaxation}
\eeq
gives 
\beq
\frac{\dr}{\dr t}x^{\,a}
=0,\qquad 
\frac{\dr}{\dr t}y_{\,a}
=\frac{\partial\,\varpi}{\partial x^{\,a}}-y_{\,a},\qquad 
\frac{\dr}{\dr t}z
=\varpi(x)-z.
\label{contact-hamilton-flow-pi}
\eeq
\item
The Legendre submanifold generated by $\varpi$, given by 
\fr{Legendre-submanifold-psi}, is an invariant manifold for 
the contact Hamiltonian vector field.
\item
Every point on $\cC\setminus\cA_{\,\varpi}$ approaches to $\cA_{\,\varpi}$ along 
an integral curve as time develops. 
Equivalently $\cA_{\,\varpi}$ is an attractor in $\cC$.  
\item 
Let $\{x(0),y(0),z(0)\}$ be a point on $\cC\setminus\cA_{\,\varpi}$.
Then the relation between the value of $h$ at $t$ and that at $t=0$ 
is given by 
\beq
h(x(t),y(t),z(t))
=\exp(-\,t)\, h(x(0),y(0),z(0)),
\label{contact-Hamiltonian-for-relaxation-exponential-speed}
\eeq
\end{enumerate}
\end{Proposition}
For completeness, 
the component expression of the contact Hamiltonian vector field
for a given contact Hamiltonian is summarized in Appendix.
Also, the explicit time-dependence of 
\fr{contact-hamilton-flow-pi} is obtained as \cite{Goto2015} 
\beqa
x^{\,a}(t)
&=&x^{\,a}(0),\non\\
y_{\,a}(t)
&=&\frac{\partial\varpi}{\partial x^{\,a}}(x(0))
+\left(\,y_{\,a}(0)-\frac{\partial\varpi}{\partial x^{\,a}}(x(0))
\,\right)\,\e^{-\,t},\non\\
z(t)
&=&\varpi(x(0))+\left(\,
z(0)-\varpi(x(0))
\,\right)\,\e^{-\,t}.
\label{contact-hamilton-flow-pi-solution}
\eeqa

As well as the case in Section\,\ref{section-geometry-equilibrium-states}, 
the details of physical meaning in the general case  are not discussed here.
However, how to use this Proposition is explained with Example as follows.
\begin{Example}
Consider the kinetic Ising model without interaction introduced in 
Example in Section\,\ref{section-solvable-heat-bath}. 
Let $(\cC,\lambda)$ be the three-dimensional contact manifold, 
where $\cC\cong\mbbR\times(-1,1)\times\mbbR$ whose canonical coordinates are 
$x,y,$ and $z$ so that $\lambda=\dr z-y\,\dr x$. 
Identify the coordinates such that  $x=\theta$, 
$y=\avg{\sigma}_{\,\theta}$,   
and $z=\Psi$.
Then, choose the contact Hamiltonian system with the contact Hamiltonian 
$h_{\,\Ising}^{\,\Psi}$ such that 
\beq
h_{\,\Ising}^{\,\Psi}(\,\theta,\avg{\sigma}_{\,\theta},\Psi\,)
=\Psi_{\,\Ising}^{\,\eq}(\theta)-\Psi.
\label{contact-hamiltonian-Ising-Psi}
\eeq
Applying Proposition\,\ref{proposition-goto-2015} to this system, 
one has that  $\cA_{\,\Psi_{\,\Ising}^{\,\eq}}$ defined 
in \fr{Ising-Legendre-submanifold-Psi} is an invariant manifold and an 
attractor.
Also, one concludes that 
$h_{\,\Ising}^{\,\Psi}(\,\theta,\avg{\sigma}_{\,\theta},\Psi\,)$ 
decreases exponentially as time develops if an initial point is 
on $\cC\setminus\cA_{\,\Psi_{\,\Ising}^{\,\eq}}$.
From the physical interpretations of $\Psi_{\,\Ising}^{\,\eq}$ and $\Psi$, 
one can interpret $h_{\,\Ising}^{\,\Psi}$ in a physical language as follows. 
The value of 
$h_{\,\Ising}^{\,\Psi}$ at time $t$ can be  interpreted as the difference between 
dimensionless free-energy at the state specified by $t$ and that at 
the equilibrium state, realized in the limit $t\to\infty$. 
This type of contact Hamiltonian will be used in 
Proposition\,\ref{fact-moment-dynamics-contact-Hamiltonian-system}.  

As well as the set of the identifications above, 
the other set of the identifications
$x=\eta$, $y=\avg{\theta}_{\,\eta}$, 
$z=\Phi$ and $h=h_{\,\Ising}^{\,\Phi}$ with 
\beq
h_{\,\Ising}^{\,\Phi}(\,\eta,\avg{\theta}_{\,\eta},\Phi\,)
=\Phi_{\,\Ising}^{\,\eq}(\eta)-\Phi,
\label{contact-hamiltonian-Ising-Phi}
\eeq
enables one to discuss relaxation processes.
Then the value of $h_{\,\Ising}^{\,\Phi}$ at time $t$ can be interpreted as the 
difference between negative entropy at time $t$ and 
that at the equilibrium state, realized in the limit $t\to\infty$. 
This type of contact Hamiltonian will be used in 
Proposition\,\ref{fact-parameter-dynamics-contact-Hamiltonian-system}. 

\end{Example} 
Some basics and 
physical applications of contact Hamiltonian systems are summarized in
Ref.\,\cite{Bravetti2017}. Also for some discussions on nonequilibrium 
processes in terms of contact geometry, see Ref.\,\cite{Bravetti2015}.

Relations among introduced spaces and manifolds are as follows. 
Let $\cP$ and $\cS_{\,\theta}, (\theta\in\Theta\subseteq\mbbR^{\,n})$ 
be the sets 
\beqa
\cP
&:=&\left\{\,p\,\bigg|\,\mbox{$p\geq 0$ can be normalized and parameterized by 
some finite set}\right\},\qquad\mbox{and}\non\\
\cS_{\,\theta}
&:=&\left\{\,p\,\bigg|\,\mbox{$p$ satisfies \fr{master-equation-solvable} and }\,
\sum_{j\in\Gamma }p(j,0;\theta)=1\,
\,\right\},
\non
\eeqa
respectively. 
Then the geometrization of $\cP$, the space of distribution functions,  
is information geometry.
In particular, dually flat space    
$(\cP,g^{\,\F},\nabla,\nabla^{\,*})$ has intensively been studied.   
Besides, the present geometry is about a contact manifold 
$(\cC,\lambda)$ provided  
maps $(\Gamma,\Theta)\to\cC$, where these maps consist of  
the integration of quantities over random variables with the weight of a 
distribution function in $\cS_{\,\theta}$, and identity maps. 
Since a target distribution function belongs 
to the exponential family, 
a convex function $\Psi^{\,\eq}$ exists. 
Then $((\cC,\lambda),\Psi^{\,\eq})$ induces 
$(\cA_{\,\Psi^{\,\eq}},g^{\,\F},\nabla,\nabla^{\,*})$, which enables 
one to deal with information geometry of equilibrium states 
from a viewpoint of contact geometry.
Also, a map $\cC\to\cC$ is realized by a flow 
of a contact Hamiltonian vector field, and is identified with a nonequilibrium 
process. As a particular case, choosing a contact Hamiltonian to be 
\fr{contact-Hamiltonian-for-relaxation} with $\varpi=\Psi^{\,\eq}$, 
one has a relaxation process. Here such a relaxation process is defined 
such that an 
integral curve connects a point of 
 $\cC\setminus\cA_{\Psi^{\,\eq}}$ and that of $\cA_{\Psi^{\,\eq}}$. 
Also, one can choose $\varpi=\Phi^{\,\eq}$ for expressing a relaxation process.

\begin{Remark}
In Ref.\,\cite{Goto2015}, a class of contact Hamiltonians was chosen 
for expressing relaxation processes. 
They are of the form  
\beq
h(x,y,z)
=\wh{h}(\varpi(x)-z),
\label{contact-Hamiltonian-for-relaxation-wider}
\eeq 
with some function $\wh{h}$. Note that this class contains 
\fr{contact-Hamiltonian-for-relaxation}, which can be verified  
by choosing $\wh{h}$ such that $\wh{h}(\Upsilon)=\Upsilon$ 
with $\Upsilon:=\varpi(x)-z$.  
In general, the relaxation rate  
for the case of \fr{contact-Hamiltonian-for-relaxation-wider}
is not exponential. On the other hand, 
the relaxation rate for the case of  \fr{contact-Hamiltonian-for-relaxation} 
 is exponential 
(See  \fr{contact-Hamiltonian-for-relaxation-exponential-speed}), 
which is the same as that for 
 \fr{master-equation-solvable-solution} and 
\fr{master-equation-solvable-solution-dual}.    
For this reason the original contact Hamiltonian 
\fr{contact-Hamiltonian-for-relaxation} is only considered in this paper.
\end{Remark}
\begin{Remark}
Divergences are defined in information geometry and they play various 
roles \cite{AN,Ay2017}.
They are often discussed on dually flat spaces in information geometry. 
 In this paper,  connections are not 
introduced in contact manifolds except for Legendre submanifolds.  
For this reason 
aspects on divergences on contact manifolds are not considered
here. However it is shown that the negative relative entropy that 
can be written as a form of divergence 
$\cS_{\,\theta}\times\cS_{\,\theta}\to\mbbR$,   
\beq
D_{\,\theta}(\,p\,\|\,p_{\,\theta}^{\,\eq})
:=\sum_{j\in\Gamma}\,p(j,t;\theta)\ln
\left[\frac{p(j,t;\theta)}{p_{\,\theta}^{\,\eq}(j)}\right], 
\label{negative-relative-entropy}
\eeq 
and the primary master equations yield 
an inequality for the solvable master equations. 
Substituting \fr{master-equation-solvable} into 
\fr{negative-relative-entropy}, one has 
$$
\frac{\dr}{\dr t}D_{\,\theta}(\,p\,\|\,p_{\,\theta}^{\,\eq})
=-\,\sum_{j\in\Gamma}(\,p(j,t;\theta)-p_{\,\theta}^{\,\eq}(j)\,)\,
(\,\ln p(j,t;\theta) -\ln p_{\,\theta}^{\,\eq}(j)\,)
\leq 0,
$$
where the inequality  
$$
(\,\varsigma-\varsigma^{\,\prime}\,)\,(\ln \varsigma - \ln\varsigma^{\,\prime})
\geq 0,\qquad\mbox{for}\quad \varsigma,\varsigma^{\,\prime}\ 
> 0  
$$
has been used. Similarly one discusses an inequality of the  
divergence depending on $p$ and $p_{\,\eta}^{\,\eq}$,
$$
D_{\,\eta}(\,p\,\|\,p_{\,\eta}^{\,\eq})
:=\sum_{j\in\Gamma}\,p(j,t;\eta)\ln
\left[\frac{p(j,t;\eta)}{p_{\,\eta}^{\,\eq}(j)}\right]. 
$$
It then follows that 
$$
\frac{\dr}{\dr t}
D_{\,\eta}(\,p\,\|\,p_{\,\eta}^{\,\eq})
\leq 0.
$$
\end{Remark}
\begin{Remark}
For nonequilibrium states, 
it is expected that flows of the solvable master equations form 
a dually flat space. 
\end{Remark}
\subsection{Geometry of primary and dual moment dynamical systems}
In this subsection Propositions\,\ref{moment-dynamics} and 
\ref{parameter-dynamics} are written in a contact geometric language.
In what follows phase space is identified with a $(2n+1)$-dimensional 
contact manifold, $(\cC,\lambda)$.

As shown below, the dynamical systems stated in these propositions 
are contact Hamiltonian systems.
\begin{Proposition}
\label{fact-moment-dynamics-contact-Hamiltonian-system}
(Primary moment dynamical system as a contact Hamiltonian system). 
The dynamical system in Proposition\,\ref{moment-dynamics}
can be written as a contact Hamiltonian system.
\end{Proposition}
\begin{Proof}
Identify $x,y,z$ and $\varpi$ in \fr{contact-hamilton-flow-pi} with   
$$
x^{\,a}
=\theta^{\,a},\qquad 
y_{\,a}
=\avg{\cO_{\,a}}_{\,\theta},\qquad
\varpi(x)
=\Psi^{\,\eq}(\theta),\qquad 
z=\Psi.
$$
One then sees that \fr{contact-hamilton-flow-pi} is identical to 
the dynamical system in Proposition\,\ref{moment-dynamics}.
This system is generated by the contact Hamiltonian 
$h$ such that 
\beq
h(\theta,\avg{\cO}_{\,\theta},\Psi)
=\Psi^{\,\eq}(\theta)-\Psi,
\label{contact-hamiltonian-Psi}
\eeq
where 
$\avg{\cO}_{\,\theta}=\{\avg{\cO_{\,1}}_{\,\theta},\ldots,\avg{\cO_{\,n}}_{\,\theta}\}$.  
\qed
\end{Proof}
Applying this Proposition
to Example in Section\,\ref{section-solvable-heat-bath},  
one has the contact Hamiltonian \fr{contact-hamiltonian-Ising-Psi}.  
Similar to this Proposition, one has the following.
\begin{Proposition}
\label{fact-parameter-dynamics-contact-Hamiltonian-system}
(Dual moment dynamical system as a contact Hamiltonian system). 
The dynamical system in Proposition\,\ref{parameter-dynamics}
can be written as a contact Hamiltonian system.
\end{Proposition}
\begin{Proof}
Identify $x,y,z$ and $\varpi$ in \fr{contact-hamilton-flow-pi} with   
$$
x^{\,a}
=\eta_{\,a},\qquad 
y_{\,a}
=\avg{\theta^{\,a}}_{\,\eta},\qquad
\varpi(x)
=\Phi^{\,\eq}(\eta),\qquad 
z=H.
$$
One then sees that \fr{contact-hamilton-flow-pi} is identical to 
the dynamical system in Proposition\,\ref{parameter-dynamics}.
This system is generated by the contact Hamiltonian 
$h$ such that 
\beq
h(\eta,\avg{\theta}_{\,\eta},\Phi)
=\Phi^{\,\eq}(\eta)-\Phi,
\label{contact-hamiltonian-Phi}
\eeq
where 
$\avg{\theta}_{\,\eta}
=\{\avg{\theta^{\,1}}_{\,\eta},\ldots,\avg{\theta^{\,n}}_{\,\eta}\}$.  
\qed
\end{Proof}
Applying this Proposition
to Example in Section\,\ref{section-solvable-heat-bath},   
one has the contact Hamiltonian \fr{contact-hamiltonian-Ising-Phi}. 

The following are geometric descriptions of the  
 Propositions\,\ref{moment-dynamics} and \ref{parameter-dynamics}.
 \begin{enumerate}
 \item
 Let $X_{\,\Psi}$ be a vector field associated with 
 the dynamical system in Proposition\,\ref{moment-dynamics}, 
 \beqa
 X_{\,\Psi}
 &:=&\frac{\dr\theta^{\,a}}{\dr t}\frac{\partial}{\partial \theta^{\,a}}
 +\frac{\dr{\avg{\cO_{\,a}}}_{\,\theta}}{\dr t}
 \frac{\partial}{\partial \avg{\cO_{\,a}}_{\,\theta}}
 +\frac{\dr\Psi}{\dr t}\frac{\partial}{\partial \Psi}
 \non\\
 &=&
 \left\{\,
 -\avg{\cO_{\,a}}_{\,\theta}+\frac{\partial\Psi^{\,\eq}}{\partial\theta^{\,a}}
 \,\right\}\frac{\partial}{\partial \avg{\cO_{\,a}}_{\,\theta}}
 +\left(\,-\Psi+\Psi^{\,\eq}\,\right)
 \frac{\partial}{\partial \Psi},
 \label{vector-field-Psi}
 \eeqa
 and $\Exp(t X_{\,\Psi}):\cC\to\cC$ 
 the exponential map associated with $X_{\,\Psi}$. 
 Then the point in $\cC$ at time $t$ is obtained by applying 
$\Exp(t X_{\,\Psi})$ 
to an  
initial point $q(0)$. That is, $q(t)=\Exp(t X_{\,\Psi}) q(0)$.
A  set of schematic pictures of this vector field on $\cC$  
is depicted in Fig.\,\ref{picture-interchange-moment}.

\begin{figure}[ht]
\begin{picture}(150,95)
\unitlength 1mm
\put(20,5){\line(1,0){43}}
\put(20,5){\line(0,1){23}}
\qbezier(22,6)(30,16)(41,16)
\qbezier(41,16)(50,16)(61,21)
\put(20,0){$0$}
\put(65,4){$\theta^{\,a}$}
\put(22,28){$\avg{\cO_{\,a}}_{\,\theta}$}
\put(64,22){$\avg{\cO_{\,a}}_{\,\theta}^{\,\eq}$}
\put(33,6){\vector(0,1){4}}
\put(33,11){\vector(0,1){3}}
\put(43,6){\vector(0,1){5}}
\put(43,12){\vector(0,1){3}}
\put(54,6){\vector(0,1){6}}
\put(54,13){\vector(0,1){4}}
\put(33,20){\vector(0,-1){3}}
\put(33,25){\vector(0,-1){4}}
\put(43,20){\vector(0,-1){3}}
\put(43,26){\vector(0,-1){4}}
\put(54,22){\vector(0,-1){3}}
\put(54,27){\vector(0,-1){3}}
\put(100,5){\line(1,0){23}}
\put(100,5){\line(0,1){23}}
\qbezier(101,13)(110,14)(121,24)
\put(100,0){$0$}
\put(125,4){$\{\theta^{\,a}\}$}
\put(102,26){$\Psi$}
\put(123,24){$\Psi^{\,\eq}(\theta)$}
\put(109,6){\vector(0,1){4}}
\put(109,11){\vector(0,1){3}}
\put(115,6){\vector(0,1){6}}
\put(115,13){\vector(0,1){4}}
\put(120,6){\vector(0,1){7}}
\put(120,14){\vector(0,1){6}}
\put(109,20){\vector(0,-1){3}}
\put(109,25){\vector(0,-1){4}}
\put(115,24){\vector(0,-1){4}}
\end{picture}
\caption{Schematic pictures of the vector field on $\cC$ for the  
primary moment dynamical system.
The attractor is the 
Legendre submanifold 
$\cA_{\,\Psi^{\,\eq}}=\{\avg{\cO_{\,a}}_{\,\theta}=\avg{\cO_{\,a}}_{\,\theta}^{\,\eq}\,\mbox{and}\,\Psi=\Psi^{\,\eq} \}$.
}
\label{picture-interchange-moment}
\end{figure}

\item
Interchange the dynamical variables with static variables in such a way that 
$$
\theta^{\,a}\mapsto
\avg{\theta^{\,a}_{\,\eta}}(t),\qquad\mbox{and}\qquad
\avg{\cO_{\,a}}_{\,\theta}(t)\mapsto\eta_{\,a},\qquad a\in\{1,\ldots,n\}.
$$
Also, $\Psi(\theta,t)\mapsto H(\eta,t)$. 
Then, let $X_{\,H}$ be a vector field associated with 
the dynamical system in Proposition\,\ref{parameter-dynamics},
\beqa
X_{\,H}
&:=&\frac{\dr\avg{\theta^{\,a}}_{\,\eta}}{\dr t}
\frac{\partial}{\partial \avg{\theta^{\,a}}_{\,\eta}}
+\frac{\dr\eta_{\,a}}{\dr t}
\frac{\partial}{\partial \eta_{\,a}}
+\frac{\dr H}{\dr t}\frac{\partial}{\partial H} 
\non\\
&=&
\left\{\,
-\avg{\theta^{\,a}}_{\,\eta}
+\frac{\partial\Phi^{\,\eq}}{\partial\eta_{\,a}}
\,\right\}
\frac{\partial}{\partial \avg{\theta^{\,a}}_{\,\eta}}
+(\,-H+\Phi^{\,\eq}\,)\frac{\partial}{\partial H} 
\label{vector-field-H}
\eeqa
and $\Exp(t X_{\,H}): \cC\to \cC$ 
the exponential map associated with $X_{\,H}$. 
Then the point in $\cC$ at time $t'+t$  is 
obtained by applying $\Exp(t' X_{\,H})$ to $q(t)$. That is,  
$q(t'+t)=\Exp(t' X_{\,H}) q(t)$.
A set of schematic pictures of this vector field on $\cC$  
is depicted 
in Fig.\,\ref{picture-interchange-parameter}.

\begin{figure}[ht]
\begin{picture}(150,95)
\unitlength 1mm
\put(20,5){\line(1,0){43}}
\put(20,5){\line(0,1){23}}
\qbezier(22,6)(30,16)(41,16)
\qbezier(41,16)(50,16)(61,21)
\put(20,0){$0$}
\put(65,4){$\eta_{\,a}$}
\put(22,28){$\avg{\theta^{\,a}}_{\,\eta}$}
\put(33,6){\vector(0,1){4}}
\put(33,11){\vector(0,1){3}}
\put(43,6){\vector(0,1){5}}
\put(43,12){\vector(0,1){3}}
\put(54,6){\vector(0,1){6}}
\put(54,13){\vector(0,1){4}}
\put(64,22){$\avg{\theta^{\,a}}_{\,\theta}^{\,\eq}$}
\put(33,20){\vector(0,-1){3}}
\put(33,25){\vector(0,-1){4}}
\put(43,20){\vector(0,-1){3}}
\put(43,26){\vector(0,-1){4}}
\put(54,22){\vector(0,-1){3}}
\put(54,27){\vector(0,-1){3}}

\put(100,5){\line(1,0){23}}
\put(100,5){\line(0,1){23}}
\qbezier(101,13)(110,14)(121,24)
\put(100,0){$0$}
\put(125,4){$\{\eta_{\,a}\}$}
\put(102,26){$H$}
\put(123,24){$\Phi^{\,\eq}(\eta)$}
\put(109,6){\vector(0,1){4}}
\put(109,11){\vector(0,1){3}}
\put(115,6){\vector(0,1){6}}
\put(115,13){\vector(0,1){4}}
\put(120,6){\vector(0,1){7}}
\put(120,14){\vector(0,1){6}}
\put(109,20){\vector(0,-1){3}}
\put(109,25){\vector(0,-1){4}}
\put(115,24){\vector(0,-1){4}}
\end{picture}
\caption{Schematic pictures of the vector field on $\cC$ for the  
dual moment dynamical system.
The attractor is the 
Legendre submanifold 
$\cA_{\,\Phi^{\,\eq}}=\{\avg{\theta^{\,a}}_{\,\eta}=\avg{\theta^{\,a}}_{\,\eta}^{\,\eq}\,\mbox{and}\,H=\Phi^{\,\eq} \}$.}
\label{picture-interchange-parameter}
\end{figure}

\end{enumerate} 
 
\subsection{Curve  length from the equilibrium state }
In nonequilibrium statistical physics,  
attention is often concentrated on 
how far a state is close to a given target state.  
In general, to formulate such a distance in terms of geometric language,     
length of a curve can be used. A length is a measure 
for expressing how far given two points are away, where such points are  
connected with an integral curve of a vector field on a manifold 
equipped  
with a metric tensor field. 

Since contact manifolds have been introduced, 
one is interested in introducing a metric tensor field on contact manifolds
for defining a length.
First, a general theory is briefly summarized. Then, 
after a metric tensor field on a contact manifold is given,  
lengths between 
nonequilibrium states and target states are calculated where   
contact Hamiltonian vector fields express 
 nonequilibrium processes. 

Let $\cM$ be a manifold, $\gamma:\mbbR\to\cM$, $(t\mapsto \gamma(t))$ 
a curve on $\cM$,  
$\dot{\gamma}$ a vector field on $\cM$ that is 
obtained as the push-forward 
$\gamma_{\,*}(\partial/\partial t)$,  
and $g$ a Riemannian metric tensor 
field. Then the length of $\gamma$ for $t_{\,0}\leq t\leq t_{\,1}$ 
is defined by \cite{Nakahara1990}  
\beq
l\,[\dot{\gamma}]_{\,t_{\,0}}^{\,t_{\,1}}
=\int_{t_{\,0}}^{\,t_{\,1}}\sqrt{g(\dot{\gamma},\dot{\gamma})}\,\dr t.
\label{length-general-metric}
\eeq   
With this introduced metric tensor field, one can calculate 
the length between a state and the target state, where a contact 
Hamiltonian vector field is identified with $\dot\gamma$ in
\fr{length-general-metric}.
This can be used to estimate how far states are away 
from the equilibrium state or a target state. 

In Ref.\,\cite{Bravetti2015JPhysA}
a metric tensor field on contact manifolds has been studied.
Let $(\cC,\lambda)$ be a $(2n+1)$-dimensional contact manifold, $(x,y,z)$ 
coordinates such that $\lambda=\dr z- y_{\,a}\dr x^{\,a}$ with
$x=\{x^{\,1},\ldots,x^{\,n}\}$ and $y=\{y_{\,1},\ldots,y_{\,n}\}$. 
One defines 
\beq
G=\frac{1}{2}\left[\,
\dr x^{\,a}\otimes \dr y_{\,a}+\dr y_{\,a}\otimes \dr x^{\,a}\,
\right]+\lambda\otimes\lambda.
\label{metric-G-Bravetti}
\eeq
This metric tensor field is consistent with the one at  
the equilibrium state,
in the sense that the 
Fisher metric tensor field on a Legendre submanifold generated by 
a function is obtained with  the pull-back of $G$. 

The following gives the length along the introduced contact 
Hamiltonian vector field. 
\begin{Lemma}
\label{fact-length-Psi}
The length between a state and the equilibrium state for
the primary moment dynamical system calculated  with 
\fr{metric-G-Bravetti} is 
\beq
l[X_{\,\Psi}]_{\,\infty}^{\,t}
=|\,\Psi(\theta(t))-\Psi^{\,\eq}\,|
=|\,h(\theta,\avg{\cO}_{\,\theta},\Psi)\,|,
\label{length-Psi-h}
\eeq
where this $h$ is specified as \fr{contact-hamiltonian-Psi}.
Then the convergence rate for \fr{length-Psi-h} is exponential. 
\end{Lemma}
\begin{Proof}
Since a state is described by $X_{\,\Psi}$ in 
\fr{vector-field-Psi}. Substituting $X_{\,\Psi}$ 
into \fr{length-general-metric}, one has 
$$
l[X_{\,\Psi}]_{\,\infty}^{\,t}
=\int_{\infty}^{t}\sqrt{G(X_{\,\Psi},X_{\,\Psi})}\,\dr t^{\,\prime}
=\int_{\infty}^{t}\sqrt{\left(\frac{\dr\Psi}{\dr t^{\,\prime}}\right)^2}
\,\dr t^{\,\prime}
=|\,\Psi(t)-\Psi^{\,\eq}\,|. 
$$
With \fr{contact-hamiltonian-Psi}, the last equality above 
can be written in terms of the contact Hamiltonian $h$ as 
\fr{length-Psi-h}. 
It follows from   
\fr{contact-Hamiltonian-for-relaxation-exponential-speed} that  
the convergence rate is exponential.  
\qed
\end{Proof}
Similarly one has the following.
\begin{Lemma}
\label{fact-length-Phi}
The length between a state and the equilibrium state for
the dual moment dynamical system calculated  with 
\fr{metric-G-Bravetti} is 
\beq
l[X_{\,H}]_{\,\infty}^{\,t}
=|\,\Phi(\eta(t))-\Phi^{\,\eq}\,| 
=|\,h(\eta,\avg{\theta}_{\,\eta},\Phi)\,|, 
\label{length-Phi-h}
\eeq
where this $h$ is specified as \fr{contact-hamiltonian-Phi}.
Then the convergence rate for \fr{length-Phi-h}
is exponential.  
\end{Lemma}
\begin{Proof}
A way to prove this is analogous to the proof of 
Lemma\,\ref{fact-length-Psi}.
\qed
\end{Proof}

Combining Lemmas\,\ref{fact-length-Psi}  \ref{fact-length-Phi},  and 
discussions in the previous sections, 
one has
the main theorem in this paper.
\begin{Thm}
(Contact geometric description of the expectation variables and 
its convergence). 
Dynamical systems derived from solvable master equations are 
described on a contact manifold, and 
its convergence rate of the 
associated with 
the metric tensor field \fr{metric-G-Bravetti} is exponential. 
\end{Thm}
\section{Conclusions}
\label{section-conclusion}
This paper offers a viewpoint  
that sets of dynamical systems exactly derived from 
master equations  
can be described 
in terms of contact geometry and information geometry. 
To show this explicitly,  
a solvable toy model as a set of master equations  
has been concentrated.
From this set of master equations and 
the viewpoint of duality in the sense of information geometry, 
the dual master equations have been defined.  
To give a contact geometric description 
two contact Hamiltonian vector fields have 
been introduced. 
By reviewing an existing study, how to describe attractors of such 
contact Hamiltonian vector fields  has been discussed
in terms of information geometry.   
Then, with an introduced  metric tensor field, the convergence rate 
has been shown to be exponential.  
Throughout this paper explicit expressions of geometrical objects have been 
shown by analyzing a spin model.

There are some potential future works that follow from this study. 
\begin{enumerate}
\item
One is to apply the present approach to 
various Monte Carlo methods. 
Since the present solvable master equations, the 
primary and dual master equations, do not show any  
state transition, associated MCMC methods   
belong to a special class. 
For completeness, some other master equations are of interest 
to explore how much  
the proposing geometric approach is effective. 
\item
In addition, in information geometry, 
some divergences are useful tools to study various objects. 
For this reason, as a future work, geometry related to divergences will be 
concentrated.
\item
Also, other kinetic models are expected to give physical applications and 
implications. 
\item 
It is also interesting to introduce  more geometric objects, such as 
connections and other  
metric tensor fields, to contact manifolds for analyzing 
nonequilibrium systems. 
\item 
Besides, there have been various works on geometric theories of 
MCMC methods. Thus it is interesting to see 
if there is some link between this study   
and such existing works.  
\end{enumerate}
We believe that the elucidation of these remaining questions will develop 
geometric  theory of master equations, 
that of MCMC methods,
 and nonequilibrium statistical mechanics.  
\section*{Acknowledgments}
The author H.H.  would like to thank 
Prof. Hukushima for useful discussions, also he is partially supported 
by JSPS (KAKENHI) grant number 17H01793 and by JST CREST JPMJCR1761.
\appendix
\section{Component expression of contact Hamiltonian vector field}
In this Appendix, the component expression of contact Hamiltonian vector field
is given.
Let $(\cC,\lambda)$ be an $(2n+1)$-dimensional contact manifold 
with $\lambda$ being a 
contact one-form, $(x,y,z)$ a set of coordinates 
so that $\lambda=\dr z-y_{\,a}\dr x^{\,a}$ with $x=\{x^{\,1},\ldots,x^{\,n}\}$,
$y=\{y_{\,1},\ldots,y_{\,n}\}$, $t$ a time,  
and $h:\cC\to\mbbR$ a function called a contact 
Hamiltonian. Then the local component expression of the contact 
Hamiltonian vector field is 
\beq
\frac{\dr}{\dr t}x^{\,a}
=-\frac{\partial h}{\partial y_{\,a}},\qquad 
\frac{\dr}{\dr t}y_{\,a}
=\frac{\partial h}{\partial x^{\,a}}+y_{\,a}\frac{\partial h}{\partial z},\qquad
\mbox{and}\qquad
\frac{\dr}{\dr t}z 
= h-y_{\,a}\frac{\partial h}{\partial y_{\,a}}.
\label{contact-Hamiltonian-local-expression}
\eeq 
Note that there are some conventions, and then signs for this expression 
are not the same as another expression.
The set of equations 
\fr{contact-hamilton-flow-pi} can be derived by substituting 
\fr{contact-Hamiltonian-for-relaxation} into 
\fr{contact-Hamiltonian-local-expression}. 
A formulation with 
contact Hamiltonian systems is 
a candidate for analytical mechanics of 
dissipative systems \cite{Bravetti2017AnnPhys}. 
Also, some contact geometric description of information geometry was 
investigated in Ref.\,\cite{Goto2016}.


\end{document}